\documentclass{aa}  
\usepackage{color}
\definecolor{Dred}{rgb}{0.312,0.070,0.070}
\definecolor{Dblue}{rgb}{0.070,0.070,0.312}
\definecolor{Dgreen}{rgb}{0.070,0.312,0.070}
\definecolor{Db}{rgb}    {0.050,0.0,0.320}

\newcounter{note}
\let\oldmarginpar\marginpar
\renewcommand\marginpar[1]{\-\oldmarginpar[\raggedleft\footnotesize #1]{\raggedright\footnotesize #1}}

\usepackage{deluxetable}

\usepackage{graphicx}
\usepackage{txfonts}
\begin{document}

   \title{Interferometric Monitoring of Gamma-ray Bright AGNs:\\
      {Measuring the Magnetic Field Strength of 4C~$+$29.45}}

   \author{S. Kang
          \inst{1,2}
          S.-S. Lee
          \inst{1,2}
          J. Hodgson
          \inst{1,3}
          J.-C. Algaba
          \inst{4}
          J. W. Lee
          \inst{1}
          J.-Y. Kim
          \inst{5}
          J. Park
          \inst{6}
          M. Kino
          \inst{7,8}
          D. Kim
          \inst{9}
          \and
          S. Trippe
          \inst{9}
          }

   \institute{Korea Astronomy and Space Science Institute,
              776 Daedeok-daero, Yuseong-gu, Daejeon 34055, Korea\\
              \email{sslee@kasi.re.kr}
         \and
         University of Science and Technology, Korea,
217 Gajeong-ro, Yuseong-gu, Daejeon 34113, Korea
         \and
         Department of Physics and Astronomy, Sejong University, 209 Neungdong-ro, Gwangjin-gu, Seoul, South Korea
         \and
         Department of Physics, Faculty of Science, University of Malaya,
50603 Kuala Lumpur, Malaysia
         \and
         Max-Planck-Institut f$\rm{\ddot{u}}$r Radioastronomie,
Auf dem H$\rm{\ddot{u}}$gel 69, D-53121 Bonn, Germany
         \and
         Institute of Astronomy and Astrophysics, Academia Sinica, 
11F of Astronomy-Mathematics Building, AS/NTU No.1, Sec.4,  Roosevelt Rd, Taipei 10617, Taiwan, R.O.C.
         \and
         National Astronomical Observatory of Japan,
2-21-1 Osawa, Mitaka, Tokyo 181-8588, Japan
         \and
         Kogakuin University of Technology \& Engineering, Academic Support Center,
2665-1 Nakano, Hachioji, Tokyo 192-0015, Japan
         \and
         Department of Physics and Astronomy, Seoul National University,
Gwanak-gu, Seoul 08826, Korea
             }

   \date{}
   
   \titlerunning{iMOGABA:J1157+2914}
   \authorrunning{Kang et al. 2020}

 
  \abstract
   {}
   {We present the results of multi-epoch, multi-frequency monitoring of a blazar 4C~$+$29.45,
   which was regularly monitored as part of the Interferometric Monitoring of GAmma-ray Bright Active Galactic Nuclei (iMOGABA) program
   - a key science program of the Korean Very long baseline interferometry Network (KVN).}
   {Observations were conducted simultaneously at 22, 43, 86, and 129~GHz during
   the 4 years from 5 December 2012 to 28 December 2016.
   We also used additional data from 
   the 15~GHz Owens Valley Radio Observatory (OVRO) monitoring program.}
   {From the 15~GHz light curve, we estimated the variability time scales of the source during several radio flux enhancements.
   We found that the source experienced 6 radio flux enhancements with variability time scales of 9--187~days during the observing period, yielding corresponding variability Doppler factors of 9-27.
   From the multi-frequency simultaneous KVN observations, we were able to obtain accurate radio spectra of the source and hence to more precisely measure the turnover frequencies $\nu_{\rm r}$ of synchrotron self-absorbed (SSA) emission with a mean value of $\overline{\nu_{\rm r}}=28.9$~GHz.
   Using jet geometry assumptions, we estimated the size of the emitting region at the turnover frequency.
   We found that the equipartition magnetic field strength is up to two orders of magnitudes higher than the SSA magnetic field strength (0.001-0.1 G).
   This is consistent with the source being particle dominated.
   We performed a careful analysis of the systematic errors related to making these estimations.}
   {From the results, we concluded that the equipartition region locates upstream the SSA region.}


   \maketitle
%

\section{Introduction}

   Blazars are a category of Active Galactic Nuclei (AGNs) where the relativistic jet is oriented almost directly toward us~\citep{ang80}.
   Because of the small viewing angle~\citep[e.g., $\le$5 degrees,][]{jor+17},
   the emission from most of the blazars shows dynamic variations of total flux density and polarization,
   and many of these blazars are highly Doppler boosted~\citep{mad87}.
   Very long baseline interferometry (VLBI) studies of relativistic jets have revealed that the jets are highly collimated and exhibit acceleration~\citep{jor+17,hom+15}.
   Many studies have tried to explain how the jets can be accelerated and collimated~{\cite[e.g.,][]{hom+15,nak+13}}.
   One potential scenario is that the magnetic fields around the jet accelerate and collimate the jet~{\citep{beg84,kom+07}}.
   Meanwhile, shocks in jets make the magnetic fields aligned or tangled~\citep{mar+08}.
   Therefore, it is important to study the magnetic field properties around blazar jets.
   As a well-known example for such studies, M87, one of the brightest and closest AGNs, has been densely studied, including extensive investigation of its magnetic field properties~\citep{asa+12,asa+14,had+16,kimj+18b,mer+16,nak+18,ehtc+19,park+19a,park+19c}.
   
   One way to study magnetic fields is to measure the polarization of the emission.
   When relativistic electrons pass through magnetic fields,
   these electrons emit synchrotron radiations which can be linearly polarized by up to 70\%~\citep{hal+17}.
   Many studies have performed VLBI polarization observations of the jets in order to
   reveal the structure of the magnetic fields.~\citep[e.g.,][]{gab+17}.
   Also,~\citet{agu+18}
   found that the magnetic fields measured at shorter wavelengths were more ordered than at longer wavelengths 
   by comparing the fractional polarization at 86 and 230~GHz.
   However, it is rare to report the results of directly measuring the magnetic field strength from polarimetric measurements.
   
   Another way of studying magnetic fields is to measure the apparent core shifts of optically thick synchrotron self-absorption (SSA) emission at different frequencies~\citep{lob98,pus+12,lee+16a}.
   \citet{lob98} estimated the magnetic field strengths of several radio sources using the core shift. 
   In their estimation, they assumed that the region at 1~pc from the central engine is under equipartition condition. 
   Equipartition means that magnetic field and relativistic particle energies are equal.
   Under these assumptions, they derived the magnetic field strength $B_{\rm core}$ of various relativistic jets measuring strengths ranging from 0.3~G to 1.0~G.
   This method has been used by various other authors finding broadly consistent results~\citep[e.g.,][]{alg+12,kov+08,pus+12}.
   
   A way to directly measure the magnetic field strengths is to measure the turnover frequency of SSA emission,
   the size of the emission at the turnover frequency, and the flux density at the turnover frequency~\citep{mar83}.
   This method has been applied by several authors including~\citet{lee+17} that estimated the magnetic field strength 
   in the relativistic jet from a bright blazar S5~0716+714 to be $\le5$~mG.
   Similarly,~\citet{alg+18b} derived a magnetic field strength of $0.1$~mG in the flat spectrum radio quasar 1633+382 (4C~38.41) and 
   \citet{lee+20} obtained a magnetic field strength of 0.95-1.93~mG of magnetic field strength for a blazar OJ287.
   
   4C~$+$29.45 (also known as 1156+295, J1159+2914, and TON 599)
   is classified as a blazar~\citep{ver+10} {with a redshift of z=0.729}~\citep{zhao+11,wang+14}.
   {It shows }a wide time scale range of variations
   from hours to years~\citep{liu+13,wang+14, aha+07,rai+13}.
   Many previous studies have attempted to explain these variations~\citep{liu+13,wang+14, aha+07,rai+13}.
   \citet{liu+13} explained the cm-wave intra-day variations
   as being due to interstellar scintillation (ISS).
   \citet{wang+14} interpreted the long-term periodic variations from 1.7 to 7.5 yr
   as being due to pressure driven oscillations from the accretion disk.
   Moreover, the source shows strong activity in $\gamma$-rays.
   The {\it{Fermi}}-Large Area Telescope (LAT) detected high $\gamma$-ray activity of the source
   on 20 November 2015 (MJD~57346)~\citep{pri+19},
   with a daily averaged $\gamma$-ray flux of $4\times10^{-7} \rm{photons}~cm^{-2}~ s^{-1}$.
   A possible correlation between $\gamma$-ray flaring and radio variability has not been reported and will be studied in a follow-up paper.
   
   In this paper, we aim to study the magnetic field strength of 4C~$+$29.45
   using Korean VLBI Network (KVN) multi-frequency simultaneous observations and
   combining with other publicly available data.
   We introduce our observations and data reduction scheme in Section~\ref{sec2},
   and show our results and our data analysis in Section~\ref{sec3}.
   Then, we discuss our results in Section~\ref{sec4},
   and finally draw our conclusions in Section~\ref{sec5}.
   
\section{Observations and Data Reduction\label{sec2}}
\subsection{iMOGABA observations \label{sec2.1}}
   4C~$+$29.45 is a target as part of a monitoring program using the KVN.
   This program is known as Interferometric Monitoring of GAmma-ray Bright AGNs (iMOGABA) and
   has, since 5 December 2012 (MJD~56266), been conducting almost monthly observations of over 30 sources at 22, 43, 86, and 129~GHz simultaneously~\citep{lee+16b}.
   The purpose of the iMOGABA program is to study the correlation between
   radio and $\gamma$-ray light curves in $\gamma$-ray bright AGNs
   and to constrain the characteristics of the $\gamma$-ray flares originating in AGNs~\citep{alg+18a,alg+18b,hod+18,kimd+17,kimd+18,lee+17,park+19b}.
   The data presented in this paper are of the source 4C~$+$29.45
   and were conducted for about 4 years from 5 December 2012 (MJD~56266) to December 2016 (MJD~57750).
   This yielded 33 epochs, which were observed in snapshot modes consisting of 2 to 7 scans with 5-min duration
   at intervals of several hours.
   There is a mean cadence of $\sim$30 days with some gaps in time due to summer maintenance.
   We set the starting frequencies at 21.7, 43.4, 86.8 and 129.3~GHz with a bandwidth of 64 MHz,
   as an integer frequency ratio, allowing for the frequency phase transfer technique to 
   be applied allowing for increased detection sensitivity at higher frequencies~\citep{alg+15,hod+16}.
   Antenna pointing offsets were checked at each station for every scan by conducting cross-scan observations for each target source.
   Atmospheric opacities were measured every hour by conducting sky tipping measurements.
   Further detailed observation schemes are described in~\cite{lee+16b}.

\subsection{Data reduction and imaging\label{sec2.2}}

   \begin{figure*}
   \centering
   \includegraphics[width=\hsize]{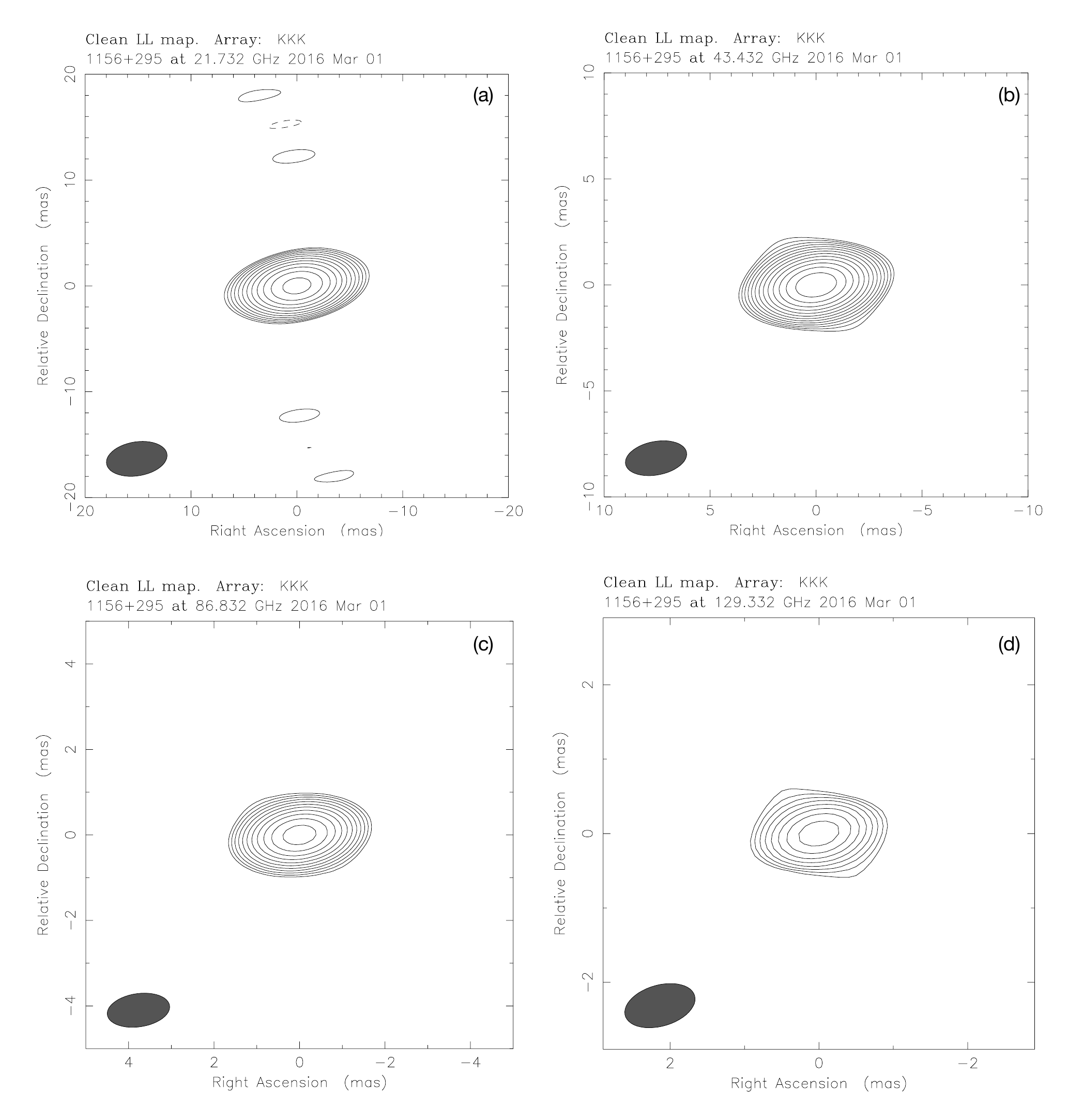}
   \caption{Example CLEANed images of 4C~$+$29.45 at 22~GHz (a), 43~GHz (b), 86~GHz (c), and 129~GHz (d), obtained on 1 March 2016. 
            The map peak fluxes are 2.39, 2.41, 1.84, and 1.25~Jy/beam at 22, 43, 86, and 129~GHz. 
            The lowest contour levels are 2.14, 1.32, 2.87, and 7.51\% of each map peak flux.
            Beam sizes are 5.79$\times$3.21, 2.93$\times$1.58, 1.47$\times$0.778, and 0.972$\times$0.552 at each frequency.}
              \label{fig1}
    \end{figure*}
   Observational data were recorded using the Mark 5B system with a recording rate of 1 Gbps.
   Recorded data at the three antennas were correlated
   with an accumulation period of 1 second,
   using the DiFX software correlator
   at the Korea-Japan Correlation Center in Daejeon, South Korea.
   Correlated data were reduced using the Astronomical Image Processing System (AIPS)
   as part of the KVN pipeline~\citep{hod+16}.
   The data reduction procedure of the KVN pipeline consists of
   amplitude calibration ({\sc APCAL}),
   correction of auto-correlations ({\sc ACCOR}),
   bandpass calibration ({\sc BPASS}),
   fringe fitting ({\sc FRING}),
   alignment of IF (sub-bands corresponding to the intermediate frequencies at the mixing step) phases using the IF-aligner~\citep{mar+12}
   and frequency phase transfer ({\sc FPT})~\citep{zhao+15}.
   The main features of the KVN pipeline that are 
   different from other pipelines are
   the use of the IF-Aligner and FPT.
   The IF-Aligner is used for calibrating delays in subbands
   and FPT is used for calibrating atmospheric errors 
   at higher frequencies with lower frequency data.
   More details about the pipeline procedure can be found in~\cite{hod+16}.
   Upper limits of the uncertainties of amplitude calibration
   were 5\% at 22 and 43~GHz, 10\%--30\% at 86 and 129~GHz.
   
   The reduced data were used for producing milliarcsecond-scale images of the source using DIFMAP.
   At the beginning of the imaging procedure, we averaged the data every 30, 30, 20, and 10 seconds
   at 22, 43, 86, and 129~GHz, respectively.
   We began by flagging outliers in amplitude.
   We then applied the CLEAN procedure with 10-100 iterations at loop gains of 0.05-0.1.
   We then applied phase self-calibration but not amplitude self-calibration due to only having three antennas~\citep{lee+14}.
   The CLEAN and self-calibration procedures were conducted until
   the residual noise is dominated by thermal random noise.
   As an example, CLEANed images at 22, 43, 86, and 129~GHz obtained on 1 March 2016 
   are shown in Fig.~\ref{fig1}.
   After the imaging procedure, we obtained
   beam parameters ($B_{\rm maj}$, $B_{\rm min}$ and $B_{\rm pa}$),
   total CLEANed flux densities ($S_{\rm KVN}$), peak flux densities ($S_{\rm p}$),
   rms noises ($\sigma_{\rm rms}$) and dynamic ranges ($D$).
   Also, we evaluated image qualities in order to verify whether the obtained images were well modeled or not,
   using the ratio between the maximum flux density and the expected flux density of the residual image, $\xi_r = s_r/s_{r,{\rm exp}}$~\citep[e.g.,][]{lee+16b}.
   The obtained parameters are listed in Table~\ref{tbl1}.
   
\subsection{Gaussian model fitting\label{sec2.3}}
   
   We fitted the source images with two-dimensional circular Gaussian models.
   We conducted this using the MODELFIT procedure in DIFMAP until the reduced chi-squares (i.e., a fitting reliability)
   did not improve significantly compared with the results of the previous fitting iterations and the residual noise became comparable to
   a uniform distribution similar to the CLEANing procedures.
   After model fitting, we obtained the best fit model parameters including the total flux density $S_{\rm total}$,
   the peak flux density $S_{\rm peak}$,
   and the size $d$ of a circular Gaussian model,
   as listed in Table~\ref{tbl2}.
   From the obtained model parameters,
   we estimated the brightness temperature $T_{\rm b}$ of core components using the following equation~\citep{lee+08},
   \begin{eqnarray}
      T_B &=& \frac{2 \rm{ln}(2)}{\pi k}\frac{S_{ {\rm tot} } \lambda^2}{d^2}(1+z)~\rm K \label{eq1},
   \end{eqnarray}
   where $k$ is the Boltzmann constant,
   $S_{{\rm tot}}$ is the total flux density in~Jy,
   $\lambda$ is the observing wavelength in cm,
   $d$ is the estimated core size in radian,
   and $z$ is redshift.
   The estimated brightness temperatures are described in Table~\ref{tbl2} and scaled by $10^{10} {\rm K}$.
   
   We derived uncertainties of each observable using the following equations~\citep{fom99,lee+08},
   \begin{eqnarray}
      \sigma_{\rm peak} & = & \sigma_{\rm rms}\left(1+ \frac{S_{\rm peak}}{\sigma_{\rm rms}}\right)^{1/2},\label{eq2}\\
      \sigma_{\rm total} & = & \sigma_{\rm peak}\left(1+ \frac{S_{\rm total}^2}{S_{\rm peak}^2}\right)^{1/2},\label{eq3}\\
      \sigma_{d} & = & d\left(\frac{\sigma_{\rm peak}}{S_{\rm peak}}\right)\label{eq4},
   \end{eqnarray}
   where these $\sigma_{\rm peak}$, $\sigma_{\rm total}$, and $\sigma_d$ are uncertainties of 
   peak flux density, total flux density, and size, respectively.
   We regarded the maximum flux and residual noise in the region corresponding to the fitted Gaussian model as $S_{\rm peak}$ and $\sigma_{\rm rms}$.
   
\subsection{Multiwavelength Data \label{sec2.4}}
   We obtained publicly available data at 15~GHz from the Owens Valley Radio Observatory (OVRO).
   The 1500 blazar monitoring program at 15~GHz with the 40 m radio telescope at OVRO began in late 2007,
   approximately one year before the start of science operations of the {\it{Fermi}}-LAT~\citep{ric+11}.
   This monitoring is a high cadence with observations conducted
   approximately twice a week and with a typical flux uncertainty of 3\%.
   4C~$+$29.45 has been included in the OVRO sample since early 2008.
   
   In addition, we used 43~GHz VLBI data from the VLBA-BU Blazar Monitoring Program (VLBA-BU-BLAZAR)\footnote{https://www.bu.edu/blazars/VLBAproject.html}. 
   This was obtained using the Very Long Baseline Array (VLBA). 
   We used this data to estimate the size of emitting regions.
   The BU blazar group has an ongoing monitoring program which observes a total of
   42 $\gamma$-ray bright blazars at 43~GHz~\citep{jorma+16}.
   In this paper, in order to estimate the sizes of emitting regions within the source,
   we obtained the calibrated visibility (uv) data from the website and fitted the data with two-dimensional circular Gaussian models in a similar way as described in Section~\ref{sec2.3}.

\section{Results and Analysis\label{sec3}}
\subsection{Multiwavelength Light Curves\label{sec3.1}}
\begin{figure*}[h!]
   \includegraphics[width=\textwidth]{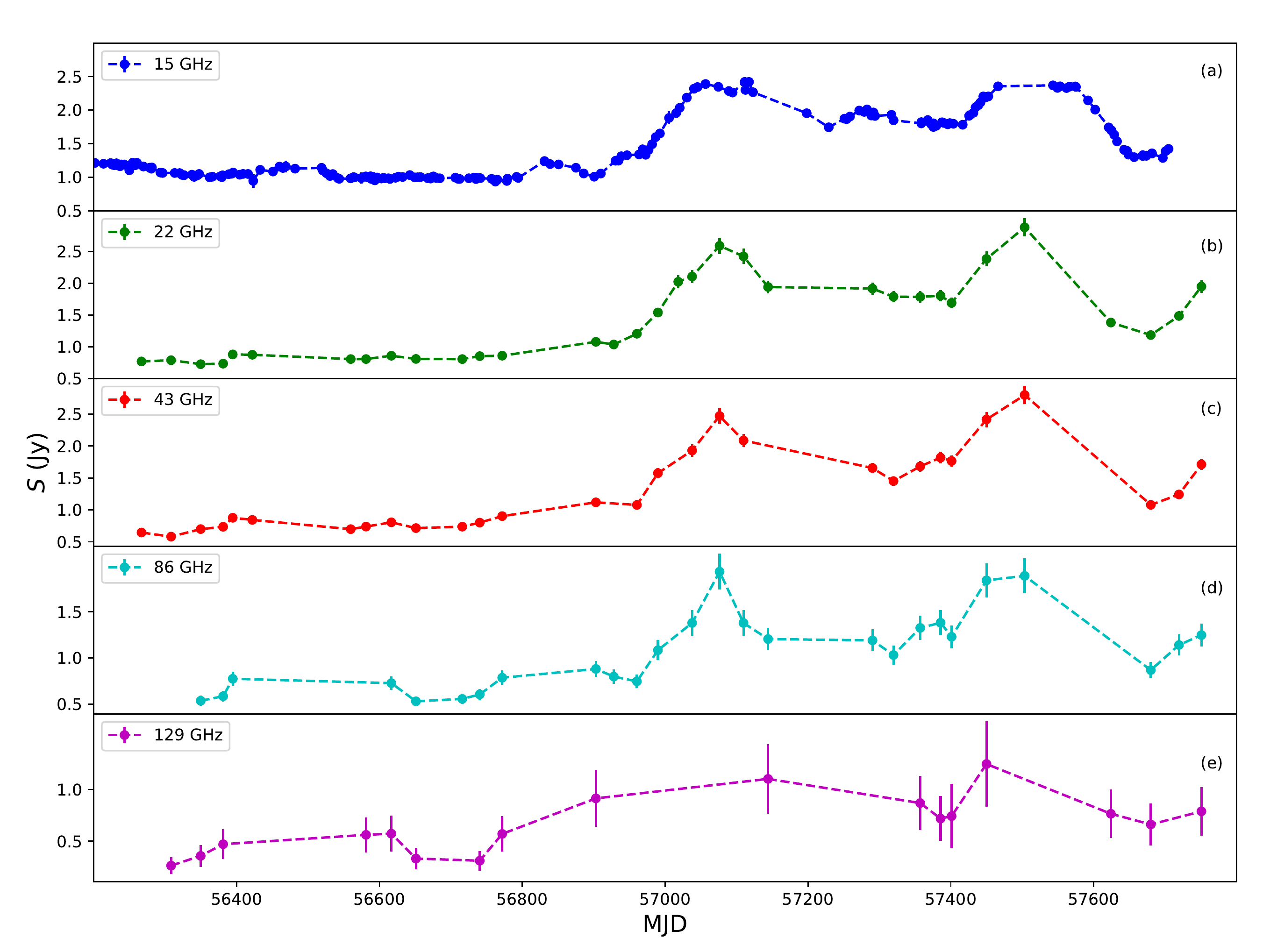}
   \centering
   \caption{Multiwavelength light curves.
   From the top to the bottom, data frequencies are 15 (obtained from OVRO),
   22, 43, 86, 129~GHz (obtained from the KVN). Dashed lines are to connect the measurements. \label{fig2}}
   \end{figure*}

   In order to study the multiwavelength behavior of the source,
   we compared the light curves at 22, 43, 86, and 129~GHz from the iMOGABA observations 
   with the OVRO (15~GHz) light curve.
   Although the 15~GHz data are obtained from a single-dish radio telescope with a large beam size
   and the 22-129~GHz data were obtained using VLBI observations with a small beam size,
   we can compare the trend of variability, assuming that the emitting region dominating the variability is compact at both scales.
   {We compared the OVRO data with VLBI flux density obtained by VLBA}~\citep{lis+18}, {and we found the mean flux ratio between VLBI and single-dish is 0.9.
   Thus we decided to combine the VLBI and single-dish data for the time-series analysis.}
   The combined multi-frequency light curves are shown in Fig.~\ref{fig2}.

\subsubsection{15~GHz Light Curve\label{sec3.1.1}}
   The 15~GHz light curve of the source is shown in the top panel of Fig.~\ref{fig2}.
   We used 211 epochs over the same period as the KVN observations
   from 5 December 2012 (MJD~56266) to 28 December 2016 (MJD~57750).
   The total flux density is in a range of 0.93-2.42~Jy during the entire observing period.
   We divided the light curve into two periods, period A and period B.
   We defined period A as before MJD~56930 and period B as after MJD~56930.
   The flux density variability of the source is quite different before and after MJD~56930:
   the source was relatively quiet from MJD~56201 to MJD~56909 (period A).
   The flux density began to increase gradually after MJD~56930 (period B).
   Period A, the source flux density is in a range of
   0.93-1.24~Jy with several minor enhancements.
   The mean flux density during the period A is 1.05~Jy with a standard deviation of 0.07~Jy.
   Period B, the source became about twice as bright
   in about 110 days after MJD~56930,
   yielding a mean flux density during the period B of 1.85~Jy with standard deviation of 0.35~Jy.
   Also, we found three flux enhancements around MJD~57100, 57300, and 57500 during the period B.
   The flux density increased 
   from 1.01~Jy to 2.42~Jy in MJD~56900-57111,
   from 1.74~Jy to 2.01~Jy in MJD~57229-57282, and
   from 1.78~Jy to 2.35~Jy in MJD~57416-57552.
   
\subsubsection{22-129~GHz Light Curves}
   The KVN 22-129~GHz light curves are shown in panels (b)-(e) of Fig.~\ref{fig2}.
   The total flux density is in a range of
   0.72-2.88~Jy,
   0.58-2.66~Jy,
   0.26-1.93~Jy, and
   0.27-1.26~Jy, at 22, 43, 86 and 129~GHz, respectively (Table~\ref{tbl1}).
   The mean flux density (and its standard deviation) is 
   1.42~Jy (0.62~Jy),
   1.28~Jy (0.59~Jy),
   0.97~Jy (0.44~Jy),
   0.64~Jy (0.26~Jy), at 22, 43, 86, and 129~GHz, respectively.
   Consistent with the OVRO light curve, we found that the KVN light curve at 22, 43, and 86~GHz can be divided by two periods.
   {The 129~GHz light curve has fewer data points due to the limitation of the KVN system and 
   we can not observe the same behavior as the lower frequencies.}
   
\subsection{Variability Time Scales\label{sec3.2}}
   We investigated the variability time scales of the source
   by decomposing the 15~GHz light curve using a number of combined exponential functions~\citep[e.g.,][]{pri+17}.
   There are several different forms of exponential equations.
   In~\cite{val99}, they divided flares into two phases, rising and decaying.
   Both phases share the peak of the flare, and the ratio between rising and decaying time scales was set to be 1.3.
   In other words, the decaying time scales are 1.3 times
   longer than their corresponding rising time scales.
   On the other hand,~\cite{pri+17} combined both rising and decaying phases into one function, without fixing the ratio of their time scales.
   We followed the method in \citet{pri+17}.

\begin{figure*}[h!]
      \includegraphics[width=\textwidth]{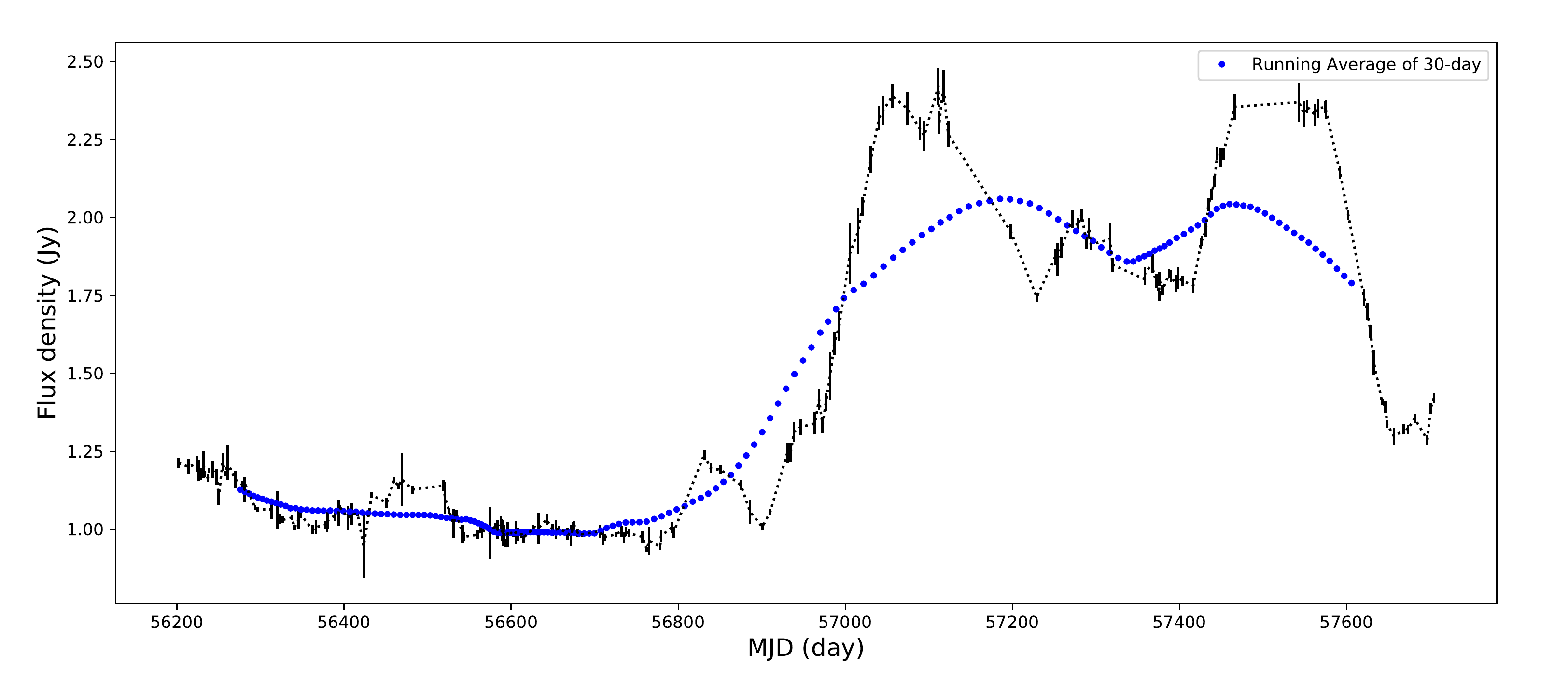}
      \centering
      \caption{30-day running average of the OVRO 15~GHz data (blue). A flare was defined as when the light curve was above the running average. 
      The times at which the flux density was above the average and the peak flux density within this time were used as initial conditions for the fitting algorithm described in Section~\ref{sec3.2}\label{fig3}.
      Dotted lines are to connect the measurements.}
   \end{figure*}

\begin{figure*}[h!]
      \includegraphics[width=\textwidth]{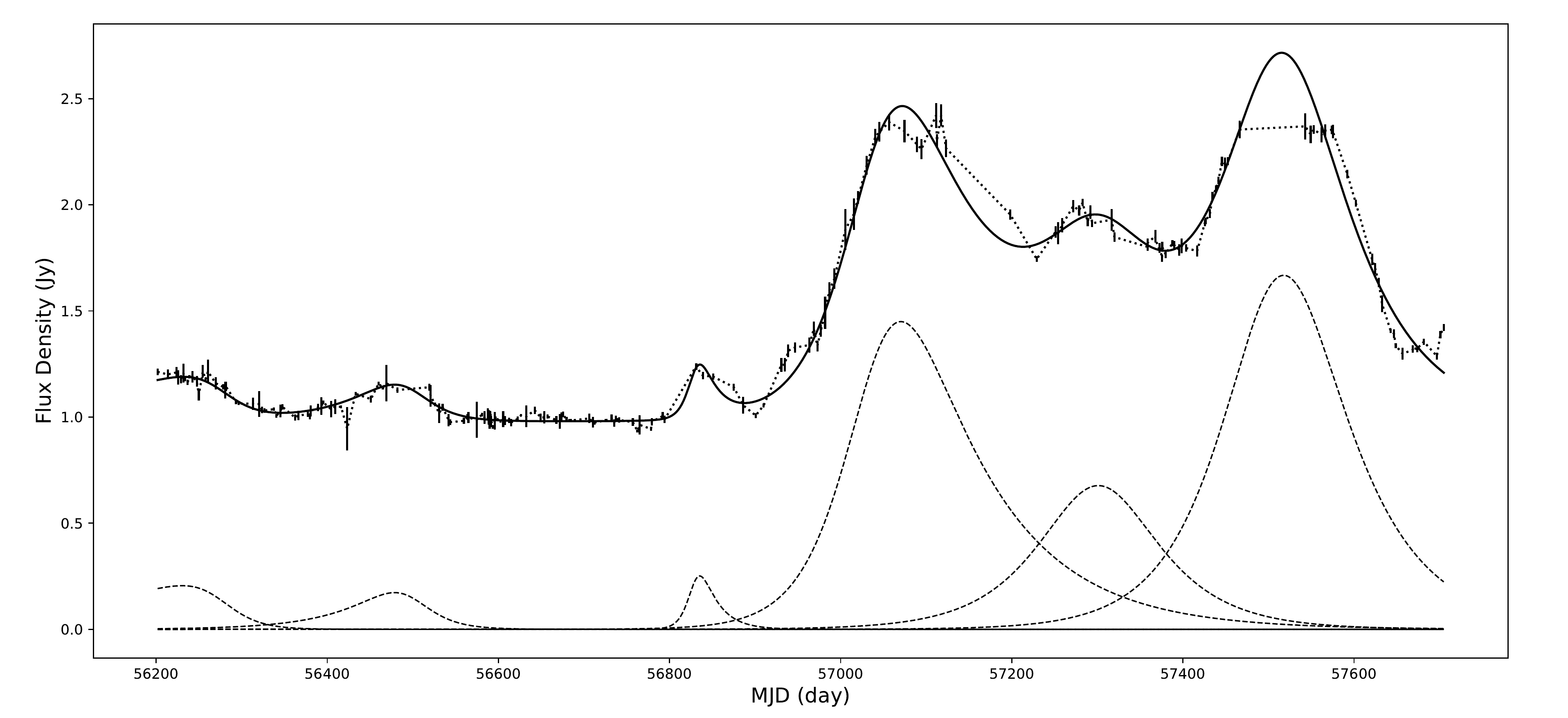}
      \centering
      \caption{Decomposition of the OVRO light curve with six flares. We could not constrain the peak for the first flare, we did not use the first flare to estimate the variability brightness temperature $T_{\rm B}^{\rm var}$ and the variability Doppler factor $\delta_{\rm var}$.
      Dotted line is to connect the measurements, dashed lines are the best fitting-models of individual flares, and solid line is the sum of the individual models.
      \label{fig4}}
   \end{figure*}

   We began by fitting the following function to the data~\citep{pri+17},
   \begin{eqnarray}
      F(t) = \sum_{i=1}^{n}2F_{0,i} \left[ {\rm exp} \left( \frac{t_{0,i} -t}{\tau_{r,i}}\right)
      + {\rm exp} \left( \frac{t-t_{0,i}}{\tau_{d,i}} \right) \right]^{-1} + b.\label{eq5}
   \end{eqnarray}
   Here, $n$ is the number of flares, and for individual flares, $F_{0,i}$ is the fitted local maximum flux density, $t_{0,i}$ is the time of the fitted maximum flux density, $\tau_{r,i}$ is the fitted rising time scale,
   $\tau_{d,i}$ is the fitted decaying time scale, and $b$ is a constant flux during our observational period.
   In order to determine the number of flares, $n$, to fit the light curve,
   we made a 30-day running average of the light curve and then defined a flare 
   as being where the flux density was above the average.
   We have plotted the running average with the light curve in Fig.~\ref{fig3}.
   The fitting algorithm requires initial conditions in order to constrain the fit.
   Where the flux density was above the average, we set the time of the local peak and the flux density of that peak 
   as the initial conditions for $t_{0,i}$ and $F_{0,i}$, respectively.
   Then we set the initial conditions for $\tau_{r,i}$ as the time between
   when the flux density began being above the average and then going below the average again.
   We found six flares ($n=6$) in this way.
   We obtained the best fitting model with a total of 25 free parameters.
   The results of this fit are summarized in Table~\ref{tbl3}.
   The results of the decomposition are shown in Fig.~\ref{fig4} and the reduced chi-squared of this fit was 8.4.
   It can be seen that the 4th flare potentially has additional structure,
   so we tried fitting with 2 functions to this flare.
   However, due to the lack of the data during the period of from MJD~57123 to 57198,
   the additional flare seems to be constructed with large uncertainty.
   Therefore, we did not fit any additional functions, even the reduced chi-squared value has been improved.
   The first flare was not well constrained and we therefore excluded it from our analysis.
   In the last flare, there was a KVN 22~GHz data point near the fitted peak from this analysis.
   The KVN data point has a value of 2.88~Jy which compares well with the fitted values in Table~\ref{tbl3},
   giving us confidence in the accuracy of the fit.
   \begin{table}[t!]
      \centering
      \caption{Best fitting parameters from the decomposition of the OVRO 15~GHz light curve.}
      \begin{tabular}{ccccc}
         \tableline
         \tableline
         &	{$F_0$}	&	{$\tau_{\rm r}$} &  {$\tau_{\rm d}$}	&	{$t_0$}		\\
         Flare	&	{(Jy)}	&	{(day)}	&	{(day)}	&	{(MJD)}		\\
         \tableline
         1	&	0.14$\pm$0.03	&	187$\pm$136 & 23$\pm$5	&	56273$\pm$11	\\
         2	&	0.16$\pm$0.01	&	62$\pm$6    & 25$\pm$4	&	56495$\pm$6	\\
         3	&	0.23$\pm$0.02	&	9$\pm$4     & 21$\pm$4	&	56830$\pm$3	\\
         4	&	1.29$\pm$0.06	&	40$\pm$2    & 118$\pm$17	&	57040$\pm$5	\\
         5	&	0.57$\pm$0.09	&	40$\pm$10   & 90$\pm$22	&	57278$\pm$12	\\
         6	&	1.61$\pm$0.04	&	57$\pm$5    & 71$\pm$1   &	57512$\pm$1	\\
         \tableline
      \end{tabular}
      \tablefoot{
      $F_0$ is the amplitude of each exponential function,
      $\tau_r$ is the rising time scale of each exponential function,
      and $t_0$ is the time of the local maximum of each exponential function.
      We obtained the best fitting model with 25 free parameters including a constant flux $b$ whose fitting result is $b=0.98$~Jy.
      \label{tbl3}}
   \end{table}
\subsubsection{Physical parameters from the variability time scales\label{sec3.2.1}}
   Once we obtained the variability time scales,
   we were able to estimate the variability brightness temperatures and
   the variability Doppler factors. 
   The Doppler factor can be estimated if there is a maximum intrinsic brightness temperature 
   ($T_{\rm B}^{\rm int}$) that the source can achieve.
   The Doppler factor is then proportional to the observed brightness temperature ($T_{\rm B}^{\rm var}$) 
   using the following equations~\citep{fuhr+08,rani+13}:
   \begin{eqnarray}
      T_{\rm B}^{\rm var} &=& 4.077\times 10^{13} \left(\frac{D_{\rm L}}{\nu~\tau_{{\rm r,}e}}\right)^2 \frac{\Delta S}{(1+z)^4} {\rm K}, \label{eq6}\\
      \delta_{\rm var} &=& \left(1+z \right) \left( \frac{T_{\rm B}^{\rm var}}{T_{\rm B, eq}} \right)^{1/3}, \label{eq7}
   \end{eqnarray}
   where $\Delta S$ is the difference of flux density between the beginning and the peak of a flare, measured in Jy, 
   $\nu$ is the observing frequency in GHz (15~GHz), $D_L$ =4477.8~Mpc (assuming $\Omega_m = 0.3$  and $H_0 = 70$~km/Mpc/s), 
   $\tau_r$ is time scale, estimated in day, and we assume that the $T_{\rm B,eq}$ is equal to $5\times 10^{10}$~K~\citep{hov+09}.
   A detailed derivation of $T_{\rm B}^{\rm var}$ is explained in Appendix~\ref{appa}.
   In this estimation, we adopt an $e$-folding time scale $\tau_{{\rm r,}e}$, which is corresponding to the time between the peak of the flare and peak/$e$.
   Moreover, we set the flux difference between peak and peak/$e$ of a flare as $\Delta S$.
   The estimated variability brightness temperatures and Doppler factors are listed in Table~\ref{tbl5}.
   \begin{table}[t!]
      \centering
      \caption{Estimated variability brightness temperatures and Doppler factors.}
      \begin{tabular}{ccc}
         \tableline
         \tableline
                 &      $T_{\rm b}^{\rm var}$  &   \\
         Flare   &      ($10^{12}$~K)  &  $\delta_{\rm var}$ \\
         \tableline
         2       &    6.52   $\pm$ 1.19     & 8.77   $\pm$  0.53  \\
         3       &    192.97 $\pm$ 78.22    & 27.12  $\pm$  3.66  \\
         4       &    47.86  $\pm$ 8.60     & 17.04  $\pm$  1.02  \\
         5       &    23.65  $\pm$ 9.23     & 13.47  $\pm$  1.75  \\
         6       &    41.85  $\pm$ 3.49     & 16.29  $\pm$  0.45  \\
         \tableline
      \end{tabular}
      \tablefoot{In the analysis, we excluded the estimation for the first flare, because it appeared only decaying phase in our period.
      \label{tbl5}}
   \end{table}
   
   In addition, we can estimate the Doppler factors from jet kinematics (e.g. apparent jet speed $\beta_{\rm app}$) assuming the critical viewing angle ($\theta=\theta_{\rm crit}$) of the jet using the following equations;
   \begin{eqnarray}
      {\rm cos}\theta_{\rm crit} &= &\beta, \\
      {\rm sin}\theta_{\rm crit} &=& \sqrt{1-\beta^2} = 1/\Gamma,  \\
      \beta_{\rm app} &=& \frac{\beta {\rm sin}\theta}{1-\beta {\rm cos}\theta} = \frac{\beta}{\sqrt{1-\beta^2}},  \\
      \delta &=& \frac{1}{\Gamma (1-\beta {\rm cos}\theta)} = \frac{1}{\Gamma(1-\beta^2)} = \Gamma =\sqrt{1+\beta_{\rm app}^2}.
   \end{eqnarray}
   In this estimation, we used the maximum apparent speed of $\beta_{\rm app} = 24.6\pm2.0$ obtained from \citet{lis+19} 
   assuming the jet viewing angle is at the critical angle, $\theta = \theta_{\rm crit}$.
   We found that the Doppler factor at the critical angle is $\delta_{\rm crit} = 24.62$,
   which is comparable with the maximum variability Doppler factor of $\delta = 27.12\pm3.66$ obtained for the flare 3, even though the observational periods are different.

\subsection{Cross-correlation Analysis\label{sec3.3}}
\begin{figure}
      \includegraphics[width=9cm]{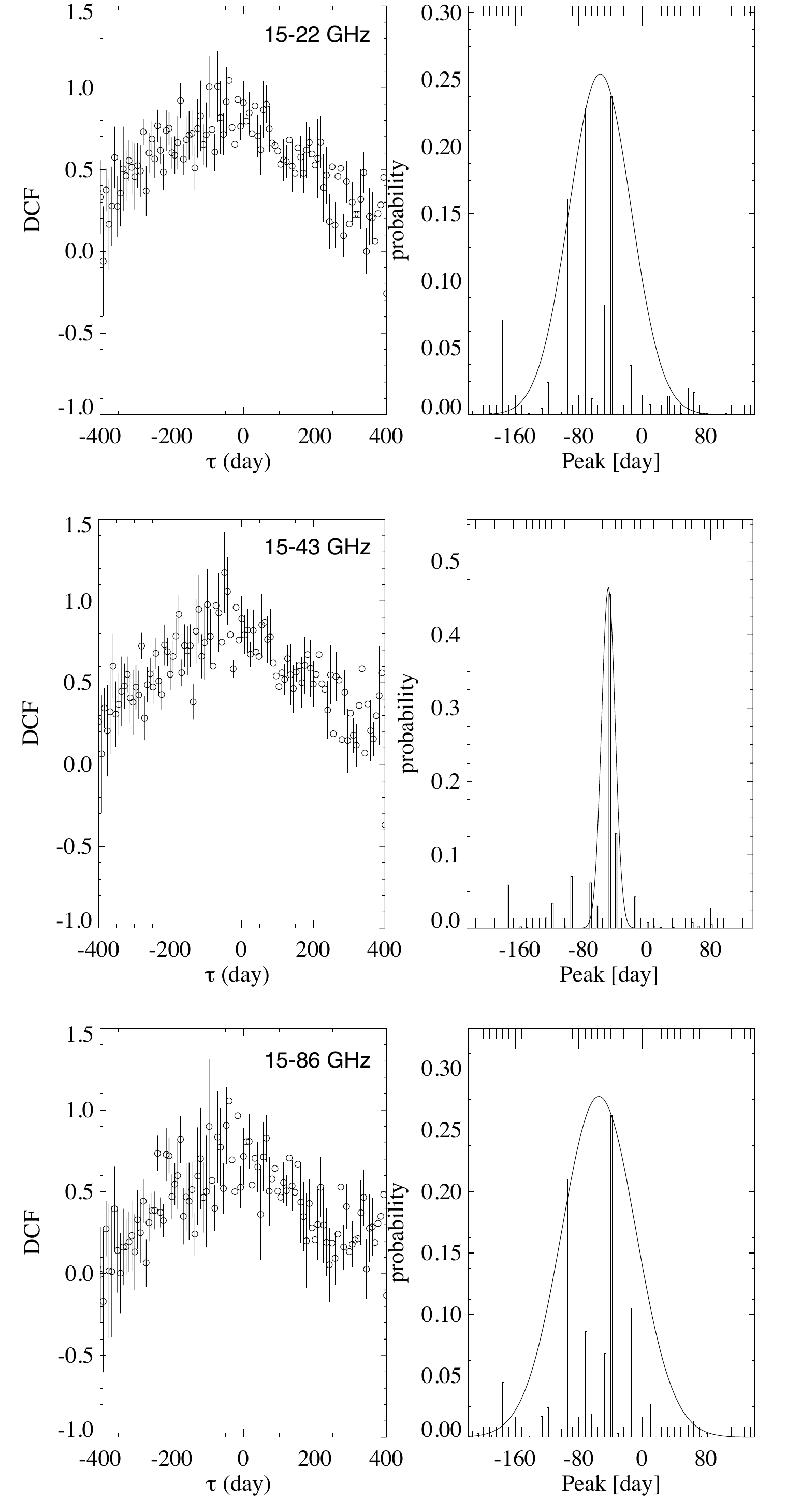}
      \centering
      \caption{Left: Samples of discrete cross-correlation functions of 15--22~GHz (top), 15--43~GHz (middle), and 
      15--86~GHz (bottom). Right: Cross-correlation peak distribution of the corresponding DCFs 
      of 1000 iterations. We fitted the distribution with a Gaussian function in order to specify the mean value and adopt the standard deviation as an uncertainty.\label{fig5}}
   \end{figure}

   In order to investigate a time lag of variations among the multi-frequency light curves,
   we compared the 15~GHz light curve with that of 22, 43, and 86~GHz,
   using a discrete cross-correlation function.
   In this analysis, we excluded the 129~GHz data
   because they are very sparse in time and have relatively large errors.
   The KVN data obtained at 22, 43, and 86~GHz were compared to the OVRO 15~GHz data in this analysis,
   yielding three data pairs as 15-22~GHz, 15-43~GHz, and 15-86~GHz.
   We used the unbinned discrete cross-correlation function (UDCF),
   in order to avoid interpolating and sampling errors~\citep{edel88}, following
   \begin{eqnarray}
      {\rm UDCF}_{ij} &=& \frac{(a_i - \bar{a})(b_j - \bar{b})}{\sqrt{(\sigma_a^2 - e_a^2)(\sigma_b^2 - e_b^2)}}, \label{eq8}\\
      {\rm DCF}(\tau) &=& \frac{1}{M} {\rm UDCF}_{ij} \label{eq9}.
   \end{eqnarray}
   Here, $a_i$ and $b_j$ are $i$th and $j$th observed data,
   $\bar{a}$ and $\bar{b}$ are mean values of the data sets $a$ and $b$,
   $\sigma_a$ and $\sigma_b$ are standard deviations of the data sets $a$ and $b$,
   and $e_a$ and $e_b$ are measurements errors, respectively. 
   The UDCF is binned with a mean cadence of $\Delta\tau$ = 8 days and 
   the number of data points $M$ in the bin, to estimate the DCF~(equation~\ref{eq9}).
   The statistical uncertainties of the time lag determined from the DCF analysis were obtained
   by performing a Monte-Carlo simulation based on a random subset selection method~\citep{pet98,lee+17}.
   Samples of DCF are shown in the left panels of Fig.~\ref{fig5}.
   A set of $N=1000$ light curves randomly sampled were used for determining the correlation peaks, 
   yielding 1000 time delays and their distribution, as shown in the right panels of Fig.~\ref{fig5}.
   By fitting the Gaussian function to the distribution, the statistical uncertainty of the time lag was determined. 
   From the discrete cross-correlation analysis, we found the time lags of
   $-54\pm35$,
   $-50\pm9$, and
   $-55\pm47$ days
   at 22, 43, and 86~GHz light curve from 15~GHz light curve, respectively.
   We also obtained the confidence interval of 95\%
   from -56 to -52,
   -51 to -49, and
   -58 to -52 at 15-22, 15-43, and 15-86~GHz, respectively~\citep{lee+17}.
   We performed this analysis for the whole period, since it is hard to distinguish the time lag of individual flares, finding that the radio flux enhancements at 22-86~GHz lead that of 15~GHz light curve.
   However it is hard to compare among 22-86~GHz, because the KVN observations were conducted simultaneously with the large cadence of about 30 days.
   Although we are not in a position to distinguish the time lag between 22 and 86~GHz light curves, the fact that the high frequency (22-86~GHz) light curves lead that of 15~GHz light curve implies an opacity effect  governing the multi-frequency (15~GHz vs 22~GHz and higher) light curve.

\subsection{Turnover Frequency\label{sec3.4}}
   Synchrotron radiation originates from relativistic electrons traveling in a magnetic field
   and has an energy distribution of $N(E)\propto E^{-\delta}$,
   where $N(E)$ is the electron number density with its energy of $E$,
   has a power-law distribution as a function of frequency, $S\propto\nu^\alpha$,
   where $\alpha$ is the spectral index.
   Theoretically, synchrotron emission from an optically thin region has the spectral index of $\alpha=(1-\delta_{e})/2$,
   where $\delta_e$ is the Doppler factor of the relativistic electrons,
   while the emission from the optically thick region has the value of $\alpha=2.5$~\citep{ryb79}.
   In order to study the characteristics of the observed synchrotron emitting region,
   we calculated the spectral indices between each pair of adjacent frequencies.
   These results are listed in Table~\ref{tbl7}, ranging from $-0.4$ to $0.1$ at 22--43~GHz, from $-0.6$ to $-0.1$ at 43--86~GHz, and 
   from $-1.6$ to $0.1$ at 86--129~GHz. We found that the source is optically thinner at higher frequencies and has a flatter spectral index at lower frequencies.
   \begin{table}[t!]
             \centering
      \caption{Spectral indices from KVN data.}
      \begin{tabular}{crrr}
         \tableline
         \tableline
         MJD        &  22-43~GHz & 43-86~GHz  & 86-129~GHz \\
         \tableline
         56266  &   $-0.25$  &       ---         &  ---   \\
         56308  &   $-0.44$  &       ---         &    ---   \\
         56349  &   $-0.05$  & $-0.38$  & $-1.02$    \\
         56381  &   $ 0.01$  & $-0.33$  & $-0.49$    \\
         56394  &   $ 0.00$  & $-0.18$  &   ---  \\
         56422  &   $-0.05$  &       ---         &   ---   \\
         56559  &   $-0.20$  &       ---         &  ---   \\
         56581  &   $-0.12$  &       ---         &   --- \\
         56616  &   $-0.09$  & $-0.15$  & $-0.59$    \\
         56651  &   $-0.17$  & $-0.43$  & $-1.11$    \\
         56716  &   $-0.13$  & $-0.41$  &  ---   \\
         56740  &   $-0.09$  & $-0.40$  & $-1.64$    \\
         56771  &   $ 0.07$  & $-0.20$  & $-0.78$    \\
         56903  &   $ 0.06$  & $-0.34$  & $ 0.11$   \\
         56960  &   $-0.16$  & $-0.53$  &   ---  \\
         56989  &   $ 0.03$  & $-0.54$  &  ---   \\
         57037  &   $-0.13$  & $-0.48$  &  ---   \\
         57076  &   $-0.07$  & $-0.35$  &  ---   \\
         57109  &   $-0.22$  & $-0.60$  &  ---   \\
         57144  &          ---        &       ---         & $-0.17$   \\
         57290  &   $-0.21$  & $-0.48$  &  ---   \\
         57319  &   $-0.30$  & $-0.49$  &  ---   \\
         57357  &   $-0.09$  & $-0.34$  & $-1.05$   \\
         57385  &   $ 0.01$  & $-0.40$  & $-1.63$   \\
         57401  &   $ 0.06$  & $-0.52$  & $-0.42$   \\
         57449  &   $ 0.02$  & $-0.39$  & $-0.74$   \\
         57503  &   $-0.04$  & $-0.57$  &  ---   \\
         57680  &   $-0.13$  & $-0.31$  & $-0.63$   \\
         57719  &   $-0.26$  & $-0.12$  &  ---   \\
         57750  &   $-0.19$  & $-0.46$  & $-1.15$   \\
         \tableline
      \end{tabular}
      \tablefoot{These spectral indices were estimated among three pairs
      from 22 to 129~GHz linearly regarding single power-law spectrum. The errors are 0.10, 0.16 and 0.79 for the pairs of 22-43, 43-86 and 86-129 GHz, respectively. 
      \label{tbl7}}
   \end{table}
   \begin{figure*}[t!]
      \includegraphics[width=\textwidth]{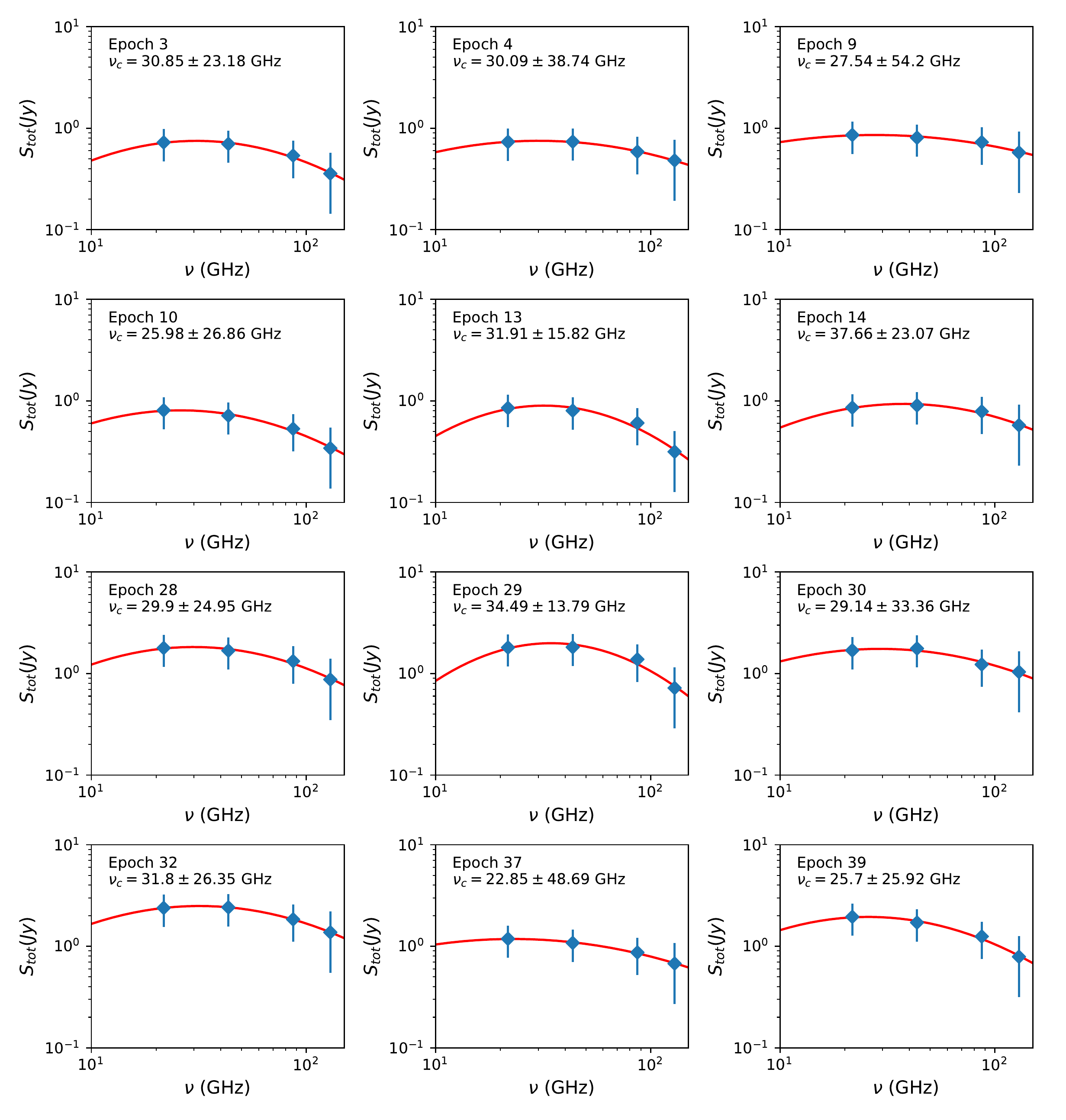}
      \centering
      \caption{The fitting results of turnover frequencies from KVN 4-frequency simultaneous data.
      The resulting turnover frequencies lie in the range of $\sim23-38$~GHz. \label{fig:fig6}}
   \end{figure*}

   Synchrotron emission can be absorbed by the same population of synchrotron particles (i.e., electrons)
   which have the same energy level at low frequencies (e.g., radio frequencies) in the optically thick region.
   This process is called SSA.
   Due to the absorption of the synchrotron radiation at lower frequencies,
   the synchrotron spectrum has a peak at a critical frequency, known as the turnover frequency, $\nu_{\rm r}$.
   This frequency is where the source transitions from being optically thick to optically thin.
   In order to determine the turnover frequency of the spectrum from the KVN multi-frequency observations,
   we fitted the observed spectra with the following curved power-law function~\citep[][and references therein]{alg+18b}:
   \begin{eqnarray}
      S(\nu) = S_{\rm r} \left(\frac{\nu}{\nu_{\rm r}} \right)^{\alpha + b {\rm ln}(\nu/\nu_r)} \label{eq10}.
   \end{eqnarray}
   Here, $S(\nu)$ is the flux density at a given frequency,
   $\nu_{\rm r}$ is a arbitrary reference frequency, $\alpha$ is the spectral index,
   $S_{\rm r}$ is the flux density for when the $\nu$ is the same as $\nu_r$,
   and $b$ is a constant.
   Because we wish to find the turnover frequency, we set $\alpha$ to be zero.
   Among the 33 epochs, we selected 13 epochs with four-frequency measurements available
   and attempted to fit the spectra.
   
   We assume that the CLEAN fluxes at four KVN frequencies originate from the core, a synchrotron self-absorbed region. However, in reality, an optically thin, extended jet can contribute to the CLEAN fluxes. The relative contribution of the jet compared with the core to the CLEAN fluxes becomes larger at lower frequencies, introducing possible bias to the data. Also, the limited uv-coverage of the KVN data may result in an artificial spectral steepening because short baselines could not be sampled at high frequencies~\citep{kimj+18a}. We calculated an artificial spectral index of the source following \cite{kimj+18a}.
   In this test, we produced simulated data to KVN array using the AIPS task {\sc UVCON} with the Stokes I image FITS data (for epoch of 2012 November 16), obtained from the MOJAVE website\footnote{http://www.physics.purdue.edu/astro/MOJAVE/allsources.html}.
   We found that the artificial spectral index of the source is 0.04, implying that the artificial spectral steepening due to the limited uv-coverage of the KVN data is negligible. We also considered additional uncertainty of 30~\%, which is obtained from a mean difference 
   between KVN CLEAN flux density and BU core flux density selected to have a time difference within 7 days at 43~GHz.
   
   We obtained turnover frequencies at 12 epochs excluding epoch 16 (MJD~56903) when the fit was failed.
   The fitting results are shown in Fig.~\ref{fig:fig6} and summarized in Table~\ref{tbl8}.
   We found that 4C~$+$29.45 has a turnover frequency ranging from 23~GHz to 38~GHz during MJD~56381--57750.
   Since the turnover frequency was measured to be below 43~GHz in all epochs when a successful fit was obtained,
   based on the curved-power-law fitting analysis. 
   we estimated the optically thin spectral index, $\alpha_{\rm thin}$,
   by performing a linear fit to the flux density at 43, 86, and 129~GHz.
   We found these to be ranging from {$-0.74$ to $-0.25$} and the results are shown in the Table~\ref{tbl8}.
   \begin{table}[t!]
      \centering
      \caption{The turnover frequency fitting results.}
      \begin{tabular}{cccc}
         \tableline
         \tableline
         MJD&   $\nu_{\rm r}$  &  $S_{\rm r}$ &  \\
         (day)    &    (GHz)    &   (Jy)    & $\alpha_{\rm thin}$  \\
         (1)    &   (2)     &   (3)     &   (4) \\
         \tableline
         56349   & 30.85$\pm$23.18   & 0.75$\pm$0.16 & -0.55$\pm$0.54 \\
         56381   & 30.09$\pm$38.74   & 0.75$\pm$0.16 & -0.37$\pm$0.56 \\
         56616   & 27.54$\pm$54.20   & 0.86$\pm$0.19 & -0.26$\pm$0.55 \\
         56651   & 25.98$\pm$26.86   & 0.81$\pm$0.19 & -0.60$\pm$0.54 \\
         56740   & 31.91$\pm$15.82   & 0.90$\pm$0.19 & -0.74$\pm$0.53 \\
         56771   & 37.66$\pm$23.07   & 0.93$\pm$0.22 & -0.35$\pm$0.54 \\
         57357   & 29.90$\pm$24.95   & 1.83$\pm$0.39 & -0.52$\pm$0.54 \\
         57385   & 34.49$\pm$13.79   & 1.99$\pm$0.44 & -0.73$\pm$0.53 \\
         57401   & 29.14$\pm$33.63   & 1.75$\pm$0.39 & -0.50$\pm$0.57 \\
         57449   & 31.80$\pm$26.35   & 2.49$\pm$0.55 & -0.48$\pm$0.55 \\
         57680   & 22.85$\pm$48.69   & 1.18$\pm$0.32 & -0.39$\pm$0.55 \\ 
         57750   & 25.70$\pm$25.92   & 1.94$\pm$0.45 & -0.64$\pm$0.54 \\
         \tableline
      \end{tabular}
      \tablefoot{(1) Modified Julian Date,
                 (2) estimated turnover frequencies,
                 (3) maximum flux densities and 
                 (4) spectral indices of optically thin region (43-129~GHz from KVN data).\label{tbl8}}
      
   \end{table}

\subsection{Magnetic Field Strength\label{sec3.5}}
   From the obtained observational parameters, we can estimate the magnetic field strength of 
   a synchrotron self-absorption region using the following equation~\citep{mar83,hod+17};
   \begin{eqnarray}
      B_{\rm SSA} = 10^{-5} b(\alpha)S_{\rm r}^{-2} \theta_{\rm r}^4 \nu_{\rm r}^5 \left( \frac{\delta}{1+z} \right)^{-1} \label{eq11}.
   \end{eqnarray}
   Here, $b(\alpha)$ is a factor depending on the spectral index~\citep[refer to Table 1 in][]{mar83}.
   In order to provide a better estimate of $b(\alpha)$ than given in~\citep{mar83}, we linearly interpolated the values listed in Table~\ref{tbl1} in~\citet{mar83},
   and fit for the optically thin spectral indices found in Section~\ref{sec3.4}.
   
   In order to estimate the size of emission region, $\theta_{\rm r}$, at the turnover frequency,
   we, first, obtained the 43~GHz images of the source from the BU program that was observed both before and 
   after the epochs where the turnover frequency could be measured.
   We then performed a simple linear interpolation in between these two VLBA epochs in order to determine the size of the 43~GHz core
   at the epochs where the turnover frequency could be measured.
   The observed core size is expected to vary as a function of frequency, following the relationship
   \begin{eqnarray}
      \theta\propto \nu^{-\epsilon},
      \label{eqtheta}
   \end{eqnarray}
   where $\epsilon$ represents the geometry of the jet.
   If the jet is conical, $\epsilon=1$, and if the jet is parabolic, $\epsilon=0.5$~\citep[e.g.,][]{alg+17}.
   In this case $\epsilon$ was found to be 0.44$\pm$0.08 by~\citet{alg+17}.
   Upper limits of the core size are obtained for 3 epochs, taking into account a minimum resolvable size~\citep[e.g.,][]{lee+16b}.
   All the uncertainties were obtained using standard error propagation method, and our derived expressions for calculating the uncertainties are given in appendix~\ref{appc}.
   Moreover, we can estimate the emission region size $\theta_{\rm r}$ from the variability size $\theta_{\rm var}$~\citep{hod+20},
   \begin{eqnarray}
      \theta_{\rm var} = 2(1+z)\frac{c\cdot\tau_{{\rm r},e}}{D_{\rm L}}\delta_{\rm var}.
   \end{eqnarray}
   In this analysis, we used the variability time scales $\tau_{{\rm r},e}$ and the variability Doppler factors $\delta_{\rm var}$ obtained in Section~\ref{sec3.2}.
   Estimated variability sizes are listed in Table~\ref{tbl10}.
   Then we extrapolated these variability sizes at turnover frequencies ($\theta_{\rm var,\nu_r}$), taking into account the jet geometry $\epsilon$, in order to compare the sizes ($\theta_{\rm VLBI,\nu_r}$) extrapolated from $\theta_{\rm VLBI}$.
   By comparing those two sizes, we found that difference between those sizes is not significant (see Section~\ref{sec4.1}).
   Therefore, we adopted the mean size between $\theta_{\rm var,\nu_r}$ and $\theta_{\rm VLBI,\nu_r}$ 
   regarding the uncertainty as the minimum and maximum of their uncertainties.
   When the variability sizes is not available we used the extrapolated core sizes $\theta_{\rm VLBI,\nu_r}$ (see Section~\ref{sec4.1}).
   Our estimated core sizes at the turnover frequency are listed in Table~\ref{tbl9}.
      
   It should be noted that for the factor $ \frac{\delta}{(1+z)}$ in equation~(\ref{eq11}), 
   we use a power index of $-1$, instead of $+1$ as in~\cite{mar83},
   because we consider that the observables originated from the core region
   which is assumed to be relatively stationary in time~\citep{lee+17,alg+18b}.
   We adopted $\delta = 9.6\pm2.6$ for the source as obtained by \citet{jor+17}.
   \begin{figure*}
      \includegraphics[width=\textwidth]{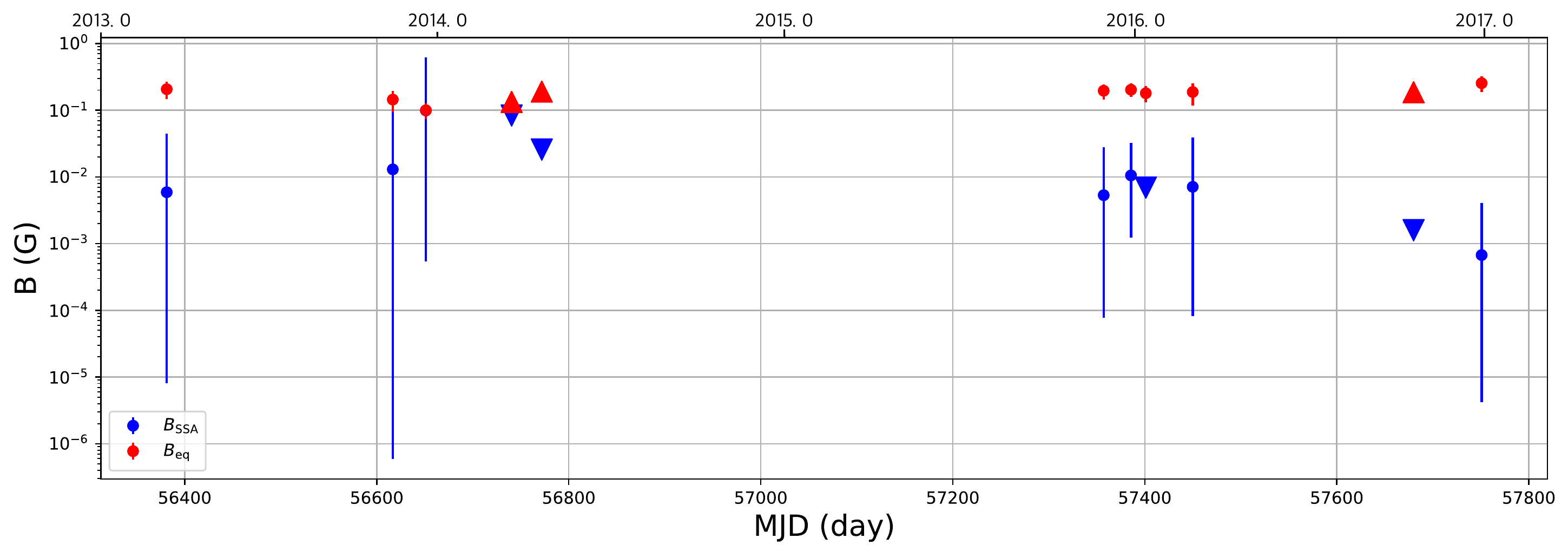}
      \centering
      \caption{Estimated magnetic field strengths of synchrotron self-absorption region and energy equipartition region. 
      Red filled circles represent $B_{\rm eq}$, while the blue filled circles represent $B_{\rm SSA}$. 
      Triangles are lower limits and inverted triangles are upper limits. 
      The lower uncertainty of $B_{\rm SSA}$ is computed in logarithm (see appendix~\ref{appc}).
      \label{fig7}}
   \end{figure*}

   Therefore, we obtained 7 measurements ranging from $0.001_{-0.001}^{+0.003}$~G to $0.099_{-0.098}^{+0.516}$~G.
   For the other epochs, only upper limits could be derived.
   In this estimation, we exclude epoch 3 (MJD~56349) because the observation was conducted during the first flare 
   which is hard to constrain the rising time scale~(\ref{sec3.2}).
   The results of this analysis are shown in Table~\ref{tbl9} and Fig.~\ref{fig7}.

   In addition to the magnetic field strength $B_{\rm SSA}$ of a synchrotron self-absorption region, 
   we can also estimate the minimum magnetic field strength $B_{\rm eq}$ 
   assuming equipartition between magnetic field and particle energy densities.
   The expression to describe this is as follows~\citep{kat+05,alg+18a}:
   \begin{eqnarray}
      B_{\rm eq} = &123\eta^{2/7} (1+z)^{11/7} \left(\frac{D_{\rm L}}{100 {\rm Mpc}}\right)^{-2/7}\left(\frac{\nu_{\rm r}}{5~{\rm GHz}}\right)^{1/7} & \nonumber \\
      \times &\left(\frac{S_{\rm r}}{100 {\rm mJy}}\right)^{2/7} \left(\frac{\theta_{\rm r}}{0.3^{\prime\prime}}\right)^{-6/7} \delta^{-5/7}\label{12}.
   \end{eqnarray}
   Here, $\eta$ is a ratio between the energies of hadrons and leptons.
   $\eta$ is 1 for a leptonic model, and $\eta$ is 1836 for a hadronic model.
   In this analysis, we adopted a compromise value of $\eta~\approx~100$
   because we do not know the true composition of the jet.
   For the same reasons as before, the equipartition magnetic field strength 
   could only be determined for 8 epochs.
   The equipartition magnetic field strengths were found to be ranging from $0.10\pm0.03$~G to $0.25\pm0.07$~G.
   These values are largely up to two orders of magnitudes stronger than $B_{\rm SSA}$, except for two epochs (MJD~56616 and MJD~56651) where $B_{\rm SSA}$ and $B_{\rm eq}$ are approximately comparable within uncertainties.
   The results are shown in Table~\ref{tbl9} and Fig.~\ref{fig7}.
   \begin{table*}
      \caption{Estimated B-Field Strengths of synchrotron self-absorption regions.}
      \centering
      \label{tbl9}
      \begin{tabular}{ccccccc}
         \hline\hline
         {MJD}&
         {$b(\alpha)$} &
         {$S_{\rm r}$} &
         {$\theta_{\rm r}$} & 
         {$\nu_{\rm r}$} & 
         {$B_{\rm SSA}$}  &  
         {$B_{{\rm eq}}$}  \\
         &                &  (Jy)         &  (mas)                    &  (GHz)           &      (G)                       &     (G)     \\
         \hline   
         56381&  2.47$\pm$3.12 & 0.75$\pm$0.16 & $0.041_{-0.004}^{+0.005}$ & 30.09$\pm$38.74  &   $0.006_{-0.006}^{+0.039}$    & $0.206_{-0.060}^{+0.059}$ \\
         56616&  1.84$\pm$3.06 & 0.86$\pm$0.19 & $0.065_{-0.005}^{+0.005}$ & 27.54$\pm$54.20  &   $0.013_{-0.013}^{+0.130}$    & $0.144_{-0.051}^{+0.051}$ \\
         56651&  3.37$\pm$0.86 & 0.81$\pm$0.19 & $0.096_{-0.007}^{+0.007}$ & 25.97$\pm$26.86  &   $0.099_{-0.098}^{+0.516}$    & $0.100_{-0.026}^{+0.026}$   \\
         56740&  3.58$\pm$0.84 & 0.90$\pm$0.19 &        $<$0.074           & 31.91$\pm$15.82  &   $<$0.084                     & $>$0.132                   \\
         56771&  2.35$\pm$3.04 & 0.93$\pm$0.22 &        $<$0.051           & 37.66$\pm$23.07  &   $<$0.026                     & $>$0.189                   \\
         57357&  3.24$\pm$0.86 & 1.83$\pm$0.39 & $0.059_{-0.005}^{+0.008}$ & 29.90$\pm$24.95  &   $0.005_{-0.005}^{+0.023}$    & $0.195_{-0.051}^{+0.048}$ \\
         57385&  3.57$\pm$0.84 & 1.99$\pm$0.44 & $0.060_{-0.008}^{+0.005}$ & 34.49$\pm$13.79  &   $0.011_{-0.009}^{+0.022}$    & $0.203_{-0.045}^{+0.048}$ \\
         57401&  3.2\tablefootmark{*}& 1.75$\pm$0.39 & $0.064_{-0.008}^{+0.006}$ & 29.14$\pm$33.36  &   $<$0.007    & $0.180_{-0.049}^{+0.051}$ \\
         57449&  3.07$\pm$3.08 & 2.49$\pm$0.55 & $0.070_{-0.021}^{+0.023}$ & 31.80$\pm$26.35  &   $0.007_{-0.007}^{+0.032}$    & $0.187_{-0.069}^{+0.066}$ \\
         57680&  2.59$\pm$3.09 & 1.18$\pm$0.32 &        $<$0.052           & 22.85$\pm$48.69  &   $<$0.002                     & $>$0.185                   \\
         57750&  3.42$\pm$0.86 & 1.95$\pm$0.45 & $0.043_{-0.003}^{+0.003}$ & 25.70$\pm$25.72  &   $0.001_{-0.001}^{+0.003}$    & $0.254_{-0.065}^{+0.065}$ \\
         \hline 
      \end{tabular}
      \tablefoot{The estimated magnetic field strength and parameters.}
   \end{table*}
   
   Magnetic field strength of a core region in the jet can be estimated from a core-shift effect as described in \citet{lob98} and \citet{pus+12}.
   Assuming equipartition between particle and magnetic field energy density, spectral index of $\alpha=-0.5$ ($S_\nu \propto \nu^{\alpha}$) and a critical angle $\theta \approx \Gamma^{-1}$, the magnetic field strength at a distance of 1~pc from the central engine is given by:
   \begin{equation}
       B_1 \simeq 0.04\Omega_{r\nu}^{3/4} (1+z)^{1/2} (1+\beta_{\rm app}^2)^{1/8}~{\rm Gauss},
   \end{equation}
   where $z$ is redshift, $\beta_{\rm app}$ is apparent jet speed and $\Omega_{r\nu}$ is the core shift measure defined in \citet{lob98} as
   \begin{equation}
       \Omega_{r\nu} = 4.85\times10^{-9} \frac{\Delta r_{c}D_{\rm L}}{(1+z)^2}\frac{\nu_1 \nu_2}{\nu_2 - \nu_1}
   \end{equation}
   where $\Delta r_c$ is the core shift between different frequencies $\nu_1$ and $\nu_2$ ($\nu_1<\nu_2$).
   The magnetic field strength $B_{\rm c}$ at a core region $r_{\rm c}$ observed by VLBI observations can be estimated as $B_{\rm c}=B_1 r_{\rm c}^{-1}$.
   The distance $r_{\rm c}$ of the core from the central engine is given by \citet{lob98} as
   \begin{equation}
       r_c(\nu) = \frac{\Omega_{r\nu}}{\nu~{\rm sin}\theta} \approx \frac{\Omega_{r\nu} (1+\beta_{\rm app}^2)^{1/8}}{\nu}~{\rm pc},
   \end{equation}
   where $\nu$ is a given frequency.
   From the study on the core shift variability of the relativistic jets~\citep{pla+19}, we found that 4C~+29.45 had a core shift of 1.02~mas measured at $\nu_1$=2.3~GHz and $\nu_2$=6.8~GHz on 2012 December 5 (MJD~56266), which is the closest measurement in time to our analysis.
   The core shift gives us a core shift measure of $\Omega_{r\nu}$=23.4~pc~GHz and magnetic field strength at 1~pc of $B_1$=1.25~G taking into account the apparent speed of the jet $\beta_{\rm app}=24.6$~\citep{lis+19}.
   In order to compare the magnetic field strength $B_{\rm c}$ from the core shift to $B_{\rm SSA}$ of synchrotron self-absorption region, we estimated the core distance $r_{\rm c}(\nu_{\rm r})$=19.14~pc at a turnover frequency of $\nu_r$=30.09~GHz (measured on MJD~56381).
   We found that the magnetic field strength $B_{\rm c}$ of the core region is $B_{\rm c}$=0.065~G which is an order of magnitude higher than $B_{\rm SSA}$=0.006~G on MJD~56381.
    
\section{Discussion\label{sec4}}
\subsection{Testing Our Assumptions on the Size Estimates\label{sec4.1}}
   According to Eq.~\ref{eq11} for estimating $B_{\rm SSA}$, the emission size $\theta_{\rm r}$ is one of the important factors for $B_{\rm SSA}$ estimates ($B_{\rm SSA} \propto \theta_{\rm r}^4$), and hence the uncertainty of the emission size.
   The determination of the emission size $\theta_{\rm r}$ presented in Section~\ref{sec3.5} rely on two assumptions.
   Firstly, the variability size (e.g., $\theta_{\rm var}$), in principle, is an intrinsic size of variable emission region (e.g., a VLBI core) and a high resolution VLBI observation may be able to resolve the emission region size (e.g., $\theta_{\rm VLBI}$).
   Therefore, at a given frequency $\nu$, those sizes are comparable ($\theta_{\rm var,\nu} \approx \theta_{\rm VLBI,\nu}$) if the VLBI observations can resolve the emission region. 
   Secondly, a jet transverse radius $R$ (or a jet angular size $\theta$) expands along the distance $r$ of a VLBI core from the central engine ($R \propto r^\epsilon$ or $\theta \propto r^\epsilon$) and the distance of the core (i.e., optical depth $\tau = 1$ surface in the jet) depends on the observing frequency $\nu$ ($r \propto \nu^{-1}$).
   Therefore, the jet geometry assumption $\theta \propto \nu^{-\epsilon}$ used to extrapolate the sizes to the turnover frequency is valid.

\begin{figure}[t!]
      \includegraphics[width=9cm]{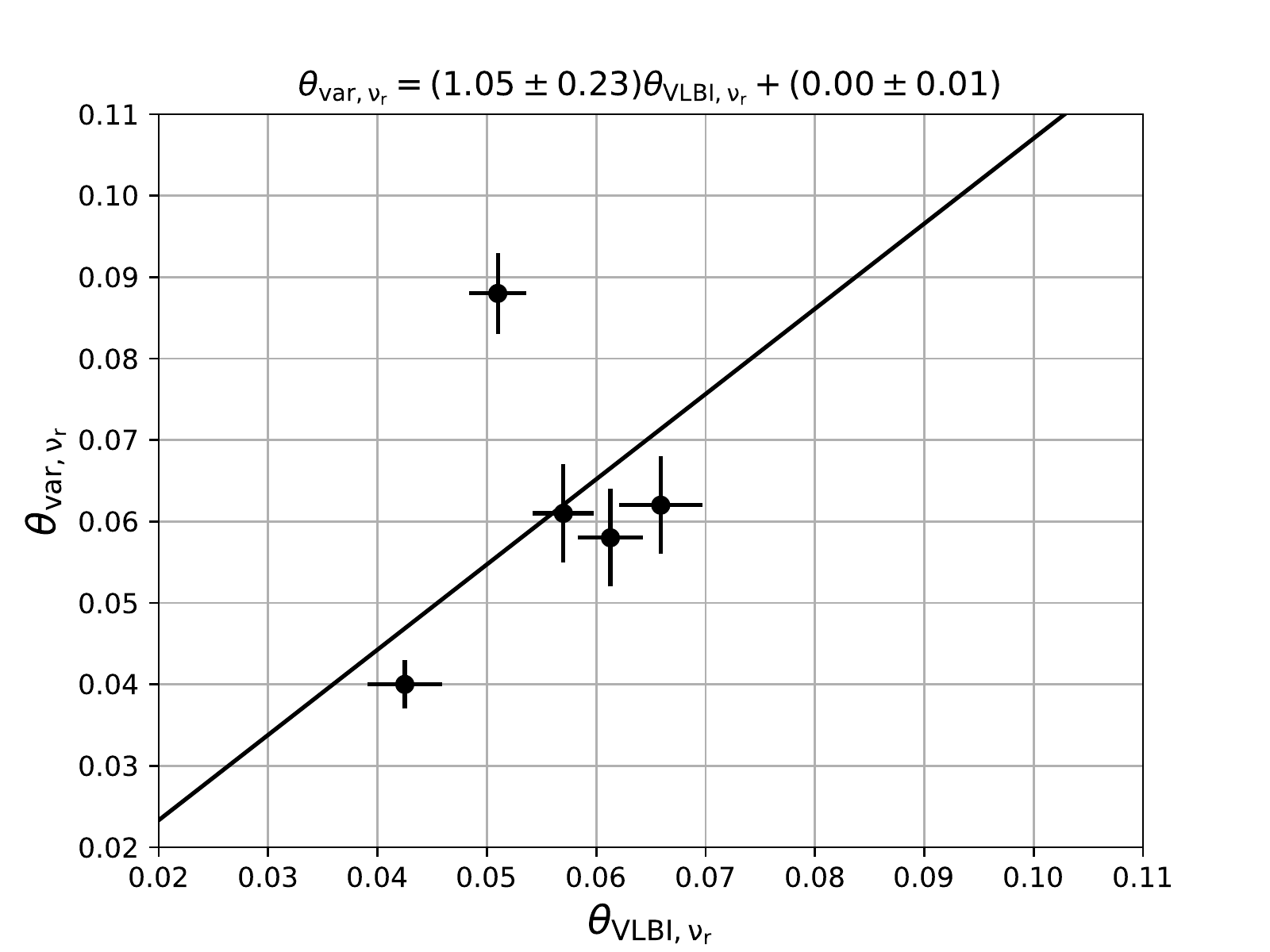}
      \centering
      \caption{Comparison between extrapolated sizes $\theta_{\rm var, \nu_r}$ and $\theta_{\rm VLBI, \nu_r}$ at turnover frequency from $\theta_{\rm var,15}$ and $\theta_{\rm VLBI,43}$ based on the jet geometry $\epsilon=0.44$~\citep{alg+17}. The solid line represents the result of the linear fitting (as also described in top of the figure).}
      \label{fig8}
   \end{figure}

   For such a case that we have a variability size $\theta_{\rm var}$ measured at a frequency $\nu_1$ and a VLBI size $\theta_{\rm VLBI}$ observed at another frequency $\nu_2$, we then expect that extrapolated sizes $\theta_{\rm var,\nu_3}$ and $\theta_{\rm VLBI,\nu_3}$ at a common frequency $\nu_3$ are comparable.
   In this study, we obtained $\theta_{\rm VLBI}$ from VLBA observations at 43 GHz and $\theta_{\rm var}$ from the variability time scales at 15 GHz.
   In order to check these assumptions, we extrapolated the sizes at turnover frequencies based on the jet geometry $\epsilon=0.44$.
   \begin{table}[h!]
      \centering
      \caption{Size comparison between VLBI measurement and variability.}
      {\small
         \begin{tabular}{ccccc}
         \tableline
         \tableline
         &  $\theta_{\rm var,15~GHz}$   & $\theta_{\rm var,\nu_r}$  & $\theta_{\rm VLBI,\nu_r}$ & $\theta_{\rm VLBI,43~GHz}$ \\
         MJD   &     $(\rm mas)$   &      $(\rm mas)$ & $(\rm mas)$   & $(\rm mas)$ \\
         (1)    &   (2) &   (3) &   (4) &   (5) \\
         \tableline
         56381 & 0.055$\pm$0.002  &  0.040$\pm$0.003 & 0.043$\pm$0.003 &  0.036$\pm$0.003  \\
         57357 & 0.083$\pm$0.007  &  0.061$\pm$0.006 & 0.057$\pm$0.003 &  0.049$\pm$0.002  \\
         57385 & 0.083$\pm$0.007  &  0.058$\pm$0.006 & 0.061$\pm$0.003 &  0.056$\pm$0.003  \\
         57401 & 0.083$\pm$0.007  &  0.062$\pm$0.006 & 0.066$\pm$0.004 &  0.056$\pm$0.003  \\
         57449 & 0.122$\pm$0.002  &  0.088$\pm$0.005 & 0.051$\pm$0.003 &  0.045$\pm$0.002  \\
         \tableline
      \end{tabular}}
      \tablefoot{(1) Date close to flare 2-6,
                 (2) variability size estimated from the variability time scale at 15~GHz,
                 (3) variability size at turnover frequency $\nu_r$ extrapolated from the variability size at 15~GHz, $\theta_{\rm var,15~GHz}$, using the jet geometry of $\epsilon=0.44$,
                 (4) the extrapolated VLBI size at turnover frequency from the VLBI size at 43~GHz, and 
                 (5) the size directly determined by VLBI observations at 43~GHz using the VLBI data from the VLBA-BU-Blazar monitoring program~\citep{jor+17}.
      \label{tbl10}}
   \end{table}
   We found that the estimated variability sizes were generally comparable with the sizes observed directly by VLBI observations as shown in Figure~\ref{fig8}.
   In order to specify the differences between $\theta_{\rm var,\nu_r}$ and $\theta_{\rm VLBI,\nu_r}$,
   we performed a linear fit to the data for 5 epochs (Fig.~\ref{fig8}).
   The fitting result shows a slope of $1.05\pm0.23$ between $\theta_{\rm var, \nu_r}$ and $\theta_{\rm VLBI, \nu_r}$, even though one data point for one epoch does not follow the trend.
   This implies that the assumptions on the size estimation are valid.
   Therefore, we decided to use the mean value of two sizes taking into the uncertainties.
   However, for the other 6 epochs, we decided to use the $\theta_{\rm VLBI, \nu_r}$, because it is hard to constrain the correlation due to the time gap between those two sizes.
   
\subsection{Brightness Temperature\label{sec4.2}}   
   We obtained the brightness temperatures from the KVN data (Table~\ref{tbl2}).
   The median brightness temperatures obtained from model fitting the KVN data are shown in Table~\ref{tbl2}.  
   We found a significant ($>3\sigma$) trend of decreasing brightness temperature with increasing frequency. 
   {In the previous studies analyzing} KVN observations, we see increasing brightness temperatures with increasing observing frequency
   because the beam size decreases with increasing observing frequency~\citep{lee+16b,alg+19}.
   Because of the large KVN beam size leading to larger core blending effect at lower frequencies,
   the KVN beam does not resolve the core at the lower frequency.
   Therefore the brightness temperature of the core was underestimated
   due to the larger source size measured.
   This effect becomes weaker at higher frequency observations which have relatively smaller beam sizes.
   In contrast to this, our results show an opposite tendency of brightness temperature to the previous studies using KVN observations,
   whereas these results are consistent to a previous study on the intrinsic brightness temperature evolution
   as function of observing frequency using high angular-resolution VLBI observations~\citep{lee14}.
   {According to}~\citet{lee14}, the intrinsic brightness temperature $T_0$ of relativistic jets decreases following 
   $T_0\propto \nu_{\rm obs}^{-1.2}$ with $\nu_{\rm obs}\ge 9~{\rm GHz}$.
   {In order to check the power index of $T_B$, we selected 14 epochs when the core sizes were estimated larger than minimum resolvable sizes at more than three frequencies.
   We found decreasing trend of core sizes along the observing frequencies in 14 epochs then fitted the $T_B$ calculated by these core sizes.
   We obtained the mean index of $-1.50\pm0.21$ (weighted average of $-1.11$), by excluding minimum resolvable sizes, 
   which is a comparable result to that estimated in}~\cite{lee14}.
   Although we are not in a position to directly derive the intrinsic brightness temperature of the relativistic jet using 
   these KVN observations, {this result makes us to conclude}
   that the KVN observations of the source are less affected by core blending effect for the 4C~$+$29.45, 
   compared with those of the previous KVN observations {for other sources}.
   
   For VLBI observations with limited visibility data, we can use a new approach for estimating brightness temperature based on individual visibility measurements and their errors following \citet{lob15}.
   We can estimate a brightness temperature limit ($T_{\rm B,lim}$) as described by equation 5 in \citet{lob15}:
   \begin{eqnarray}
   T_{\rm B,min} &=& 3.09\left(\frac{B}{{\rm km}}\right)^2 \left(\frac{V_q}{\rm mJy}\right) {\rm K},\\
   T_{\rm B,lim} &=& 1.14\left(\frac{V_q + \sigma_q}{\rm mJy}\right) \left(\frac{B}{\rm km}\right)^2 \left({\rm ln} \frac{V_q + \sigma_q}{V_q}\right) {\rm K}.
   \end{eqnarray}
   We selected KVN observations on MJD~56771, 57037, 57290 and 57503 (closest epochs to the flare 3-6), and estimated the brightness temperature limits at multi frequencies as summarized in Table~\ref{tbl11}.
   \begin{table*}
   \caption{\label{tbl11}Brightness temperatures}
   \centering
   \begin{tabular}{ccccccccccc}
   \hline\hline
   &$T_{\rm B,var}$
   &$T_{\rm B,lim,22}$&$T_{\rm B,lim,43}$&$T_{\rm B,lim,86}$&$T_{\rm B,lim,129}$
   &$T_{\rm B,22}$&$T_{\rm B,43}$&$T_{\rm B,86}$&$T_{\rm B,129}$\\
   MJD&$10^{12}$~K
   &$10^{10}$~K&$10^{10}$~K&$10^{10}$~K&$10^{10}$~K
   &$10^{10}$~K&$10^{10}$~K&$10^{10}$~K&$10^{10}$~K\\
   (1) &  (2) & (3) & (4) & (5) & (6) & (7) & (8) & (9) &(10)\\
   \hline
   56771& 192.97$\pm$78.22&3.51 &2.64 &1.05&0.47&$>$0.25&0.72$\pm$0.04&0.31$\pm$0.02&0.30$\pm$0.01\\
   57037& 47.86$\pm$8.60&16.51&11.96&2.00&--- &3.73$\pm$0.12&6.56$\pm$0.19&0.42$\pm$0.03&---  \\
   57290& 23.65$\pm$9.23&11.26&2.11 &1.04&--- &$>$1.02&1.08$\pm$0.04&0.25$\pm$0.02&---  \\
   57503& 41.85$\pm$3.49&8.67 &10.27&3.79&--- &$>$1.25&2.19$\pm$0.04&0.95$\pm$0.25&---  \\
   \hline
   \end{tabular}
   \tablefoot{(1) Epochs closest to the flare 3-6, (2) variability brightness temperatures obtained in Section~\ref{sec3.2.1}, (3)-(6) brightness temperature limits at 22, 43, 86 and 129 GHz, respectively, following \citet{lob15}, (7)-(10) brightness temperatures obtained from modelfit of KVN data at 22, 43, 86 and 129 GHz, respectively.}
   \end{table*}
   We found that the estimated brightness temperature limits ($T_{\rm B,lim}$) are comparable to those $T_{\rm B}$ obtained from KVN images, yielding the ratio of $T_{\rm B,lim}/T_{\rm B} = 1.82-4.76$.
   This is consistent to the application results of $T_{\rm B,lim}/T_{\rm B}=4.6$ for the MOJAVE data as reported in \citet{lob15}.
   
   As another way of measuring the brightness temperature of the non-thermal synchrotron emission
   from the relativistic jets, we investigated the variability brightness temperatures, $T_{\rm B}^{\rm var}$,
   from the obtained variability time scales in Table~\ref{tbl5}.
   We found that the mean variability brightness temperature is much higher than 
   the mean brightness temperature derived from the KVN images at 22~GHz and those (Table~\ref{tbl11}) obtained by applying the approach in \citet{lob15}.
   In order to understand how this difference comes, we calculated the size of the emission region,
   $\theta_{\rm var}$, from the variability properties as described in Section~\ref{sec3.5}
   and compared these $\theta_{\rm var}$ with the size of the radio core emission regions obtained from the KVN images.
   The mean core sizes are 
   $0.47\pm0.25$~mas,
   $0.53\pm0.22$~mas,
   $0.42\pm0.17$~mas, and 
   $0.37\pm0.34$~mas at 22, 43, 86, and 129~GHz, respectively.
   The obtained mean variability size is $0.088\pm0.034$~mas derived from the 15~GHz observations,
   which is smaller than the KVN size at 22~GHz by a factor of 5.
   Therefore, we postulate this discrepancy of core sizes may partially affect 
   the difference between brightness temperatures at 22-129~GHz and variability brightness temperatures at 15~GHz.
   
\subsection{Variability Behavior\label{sec4.3}}   
   We divided the radio light curve into two distinct periods: part A and B (Fig.~\ref{fig2}).
   We attempted to investigate if there is any physical changes in the source over this period.
   
   We cross-compared the flux densities with derived parameters,
   such as core sizes, turnover frequencies, and Doppler factors, yielding no cross-correlation between the parameters.
   In particular, as shown in the Tables~\ref{tbl5}, we did not find any obvious trend in the Doppler factors in section~\ref{sec3.2.1}.
   Similarly, we found no obvious trend in the variability brightness temperatures.
   This may imply that the strong flaring activities in 4C~$+$29.45 were not attributed to the Doppler factor variability.
   However, it must be noted that the Doppler factor estimates rely on an intrinsic brightness temperature assumption.
   It could be that the intrinsic brightness temperature is changing and therefore the Doppler factors as well.
   We intend to resolve this degeneracy in our future work (Kang et al. in prep.).
   
   We also compared the OVRO data with further multi-wavelength data published, including data from the {\it Fermi}-gamma-ray space telescope AGN multi-frequency monitoring alliance~\citep[F-GAMMA,][]{ang+19} and polarimetric monitoring of AGN at millimeter wavelengths~\citep[POLAMI,][]{agu+18} projects and the submillimeter array (SMA) monitoring program.
   We found two flares (flare 2 and 3) seem to appear in the F-GAMMA (23 and 43~GHz) and POLAMI (86 and 230~GHz) light curves.
   We found three flares (flare 3, 4 and 5) include the flux enhancements appeared in the SMA light curve (230~GHz).
   For the flare 2, the multi-wavelength flux densities peaked around MJD~56500 at 15, 43, 86 and 230~GHz, quasi-simultaneously.
   The spectrum seems to be inverted between 15 and 43~GHz, and steep between 43 and 230~GHz.
   On the other hand, for the flare 3, the multi-wavelength light curves at 43, 86 and 230~GHz seemed to peak earlier than the OVRO 15~GHz light curve.
   Moreover the spectrum may appear to be inverted between 15 and 86~GHz and steep between 86 and 230~GHz.
   Although we are not in a position of accurately analysing the multi-wavelengths data at 15-230~GHz, the comparison results may imply that during the period of flare 2 and 3 the inverted spectrum (optically thick spectrum) extends to higher frequency (hence higher turnover frequency) and the high energy synchrotron particles increased from the flare 2 to the flare 3 (before the source entered the period B).

\section{Conclusions}\label{sec5}   
   We studied the long-term behavior of a blazar 4C~$+$29.45 which was observed from December 2012 to December 2016.
   We analyzed KVN data with OVRO 15~GHz data.
   We found that the OVRO 15~GHz light curve can be divided into two distinguishable periods, period A and period B,
   and also found similar variation trends in 22, 43, and 86~GHz light curves.
   In order to confirm the characteristics of variation,
   we attempted to estimate the variability time scales using a combined exponential function.
   Also, we estimated turnover frequencies and maximum flux densities with 22, 43, 86, and 129~GHz KVN data.
   Then, we estimated $B_{\rm SSA}$ using turnover frequencies, maximum flux densities, with core sizes from BU data and Doppler factor of $\delta =9.6$ from \citet{jor+17}.
   The $B_{\rm SSA}$ ranged from 0.001 to 0.099~G.
   In our analysis, we used upper limits for core sizes.
   In addition, we estimated $B_{\rm eq}$ assuming the $\eta\sim100$
   that means the jet consist of electrons with both positrons and protons.
   The $B_{\rm eq}$ ranged from 0.10 to 0.25~G.
   The $B_{\rm eq}$ are much greater than the $B_{\rm SSA}$ for all periods, 
   although we assumed the upper limits and lower limits of $B_{\rm SSA}$ and $B_{\rm eq}$, respectively.
   From the results we concluded that the equipartition region locates upstream the SSA region.
   
   Meanwhile, {\it{Fermi}}-LAT reported an intense flare of gamma-ray emission from 4C~$+$29.45 in October 2015~\citep{pri+19}.
   We will discuss the correlation between gamma-ray flare and behavior of radio emission, 
   such as variation of core sizes and flux density in a follow-up study.

\begin{acknowledgements}
   We would like to thank the anonymous referee for important comments and suggestions that have enormously improved the manuscript.
   J.-C. Algaba acknowledges support from the Malaysian Fundamental Research Grant Scheme (FRGS)FRGS/1/2019/STG02/UM/02/6.   
   This work was supported by the National Research Foundation of Korea (NRF) grant funded by the Korea government (MIST) (2020R1A2C2009003).
   We are grateful to the staff of the KVN who helped to operate the array and to correlate the data. 
   The KVN and a high-performance computing cluster are facilities operated by the KASI (Korea Astronomy and Space Science Institute). 
   The KVN observations and correlations are supported through the high-speed network connections among the KVN sites provided by 
   the KREONET (Korea Research Environment Open NETwork), which is managed and operated by the KISTI (Korea Institute of Science and Technology Information).
   This research has made use of data from the OVRO 40-m monitoring program (Richards, J. L. et al. 2011, ApJS, 194, 29) 
   which is supported in part by NASA grants NNX08AW31G, NNX11A043G, and NNX14AQ89G and NSF grants AST-0808050 and AST-1109911.
   This study makes use of 43~GHz VLBA data from the VLBA-BU Blazar Monitoring Program 
   (VLBA-BU-BLAZAR\footnote{http://www.bu.edu/blazars/VLBAproject.html}), funded by NASA through the {\it{Fermi}} Guest Investigator Program. 
   The VLBA is an instrument of the National Radio Astronomy Observatory. 
   The National Radio Astronomy Observatory is a facility of the National Science Foundation operated by Associated Universities, Inc.
   This research has made use of data from the MOJAVE database that is maintained by the MOJAVE team~\citep{lis+18}.
   J.P. acknowledges financial support from the Korean National Research Foundation (NRF) via Global PhD Fellowship grant 2014H1A2A1018695 and support through the EACOA Fellowship awarded by the East Asia Core Observatories Association, which consists of the Academia Sinica Institute of Astronomy and Astrophysics, the National Astronomical Observatory of Japan, Center for Astronomical Mega-Science, Chinese  Academy of Sciences, and the Korea Astronomy and Space Science Institute.
\end{acknowledgements}

%
%

\begin{appendix}
\section{Image and Modelfit Parameters from KVN Observations.\label{appa}}
The image and modelfit parameters obtained from KVN observations, as described in Section~\ref{sec2}, are summarized in Table~\ref{tbl1} and \ref{tbl2}.
\longtab[1]{
\begin{longtable}{cccccccccccc}
\caption{Image Parameters from KVN Observations\label{tbl1}}\\
\hline\hline
&
&
&
&
{$B_{\textup{max}}$}&
{$B_{\textup{min}}$} & 
{$B_{\textup{pa}}$}  &
{$S_{\textup{KVN}}$} & 
{$S_{\textup{p}}$} &
{$\sigma_{\rm rms}$}  &
 &
 \\
Date   & MJD & Band &   & (mas) & (mas)& ($^\circ)$ &(Jy) & (Jy/beam) &(mJy/beam) & $D$ & $\xi_{\textup{r}}$\\  
(1)&(2)&(3)&&(4)&(5)&(6)&(7)&(8)&(9)&(10)&(11)\\
\hline
\endfirsthead
\caption{continued.}\\
\hline\hline
&
&
&
&
{$B_{\textup{max}}$}&
{$B_{\textup{min}}$} & 
{$B_{\textup{pa}}$}  &
{$S_{\textup{KVN}}$} & 
{$S_{\textup{p}}$} &
{$\sigma_{\rm rms}$}  &
 &
 \\
Date   & MJD & Band &   & (mas) & (mas)& ($^\circ)$ &(Jy) & (Jy/beam) &(mJy/beam) & $D$ & $\xi_{\textup{r}}$\\  
(1)&(2)&(3)&&(4)&(5)&(6)&(7)&(8)&(9)&(10)&(11)\\
\hline
\endhead         
\hline
\endfoot
         2012-12-05 & 56266 & K &  & 6.06 & 3.07 & 72.33  & 0.77 & 0.77 & 6.66  & 115.06 & 0.45 \\
                    &          & Q &  & 2.93 & 1.68 & 63.5   & 0.65 & 0.64 & 34.31 & 18.63  & 0.48 \\
         2013-01-16 & 56308 & K &  & 8.61 & 3.08 & $-$58.75 & 0.79 & 0.78 & 4.77  & 164.23 & 0.40 \\
                    &          & Q &  & 4.24 & 1.55 & $-$56.28 & 0.58 & 0.58 & 14.71 & 39.15  & 0.47 \\
                    &          & D &  & 2.23 & 0.74 & $-$55.33 & 0.27 & 0.26 & 13.52 & 19.54  & 0.50 \\
         2013-02-27 & 56349 & K &  & 6.20 & 2.99 & 76.09  & 0.72 & 0.72 & 10.18 & 70.91  & 0.45 \\
                    &          & Q &  & 3.27 & 1.43 & 77.16  & 0.70 & 0.70 & 20.88 & 33.29  & 0.47 \\
                    &          & W &  & 1.65 & 0.71 & 76.09  & 0.54 & 0.52 & 27.81 & 18.87  & 0.47 \\
                    &          & D &  & 1.12 & 0.47 & 76.4`4  & 0.36 & 0.36 & 17.20 & 20.82  & 0.56 \\
         2013-03-28 & 56381 & K &  & 5.14 & 3.18 & $-$88.34 & 0.73 & 0.73 & 5.01  & 145.85 & 0.56 \\
                    &          & Q &  & 2.56 & 1.68 & $-$82.98 & 0.74 & 0.73 & 15.50 & 47.11  & 0.56 \\
                    &          & W &  & 1.30 & 0.76 & 88.19  & 0.59 & 0.56 & 40.39 & 13.94  & 0.58 \\
                    &          & D &  & 1.15 & 0.46 & 73.25  & 0.48 & 0.47 & 54.91 & 8.58   & 0.46 \\
         2013-04-11 & 56394 & K &  & 5.47 & 3.17 & $-$74.74 & 0.88 & 0.88 & 4.78  & 183.31 & 0.52 \\
                    &          & Q &  & 2.73 & 1.64 & $-$73.65 & 0.88 & 0.87 & 13.21 & 66.10  & 0.65 \\
                    &          & W &  & 1.34 & 0.77 & $-$80.12 & 0.78 & 0.74 & 17.63 & 43.98  & 0.57 \\
         2013-05-08 & 56422 & K &  & 5.37 & 3.31 & $-$71.12 & 0.87 & 0.87 & 7.62  & 113.77 & 0.48 \\
                    &          & Q &  & 2.60 & 1.78 & $-$68.92 & 0.84 & 0.84 & 19.33 & 43.36  & 0.60 \\
         2013-09-24 & 56559 & K &  & 5.32 & 3.18 & $-$84.24 & 0.80 & 0.80 & 18.41 & 43.56  & 0.64 \\
                    &          & Q &  & 3.91 & 1.56 & $-$65.89 & 0.70 & 0.70 & 7.97  & 87.58  & 0.47 \\
         2013-10-15 & 56581 & K &  & 5.38 & 3.17 & $-$83.06 & 0.81 & 0.80 & 9.28  & 86.54  & 0.58 \\
                    &          & Q &  & 2.79 & 1.57 & $-$88.83 & 0.74 & 0.74 & 16.57 & 44.51  & 0.51 \\
                    &          & D &  & 1.35 & 0.75 & $-$81.7  & 0.57 & 0.56 & 42.31 & 13.28  & 0.60 \\
         2013-11-20 & 56616 & K &  & 6.17 & 2.99 & 77.19  & 0.86 & 0.85 & 18.16 & 46.97  & 0.43 \\
                    &          & Q &  & 3.19 & 1.45 & 75.26  & 0.81 & 0.80 & 18.86 & 42.53  & 0.48 \\
                    &          & W &  & 1.64 & 0.71 & 75.51  & 0.73 & 0.72 & 21.47 & 33.32  & 0.49 \\
                    &          & D &  & 1.05 & 0.50 & 79.88  & 0.58 & 0.57 & 20.96 & 27.34  & 0.54 \\
         2013-12-24 & 56651 & K &  & 6.62 & 2.91 & $-$61.13 & 0.81 & 0.80 & 5.79  & 138.97 & 0.47 \\
                    &          & Q &  & 3.16 & 1.53 & $-$60.41 & 0.72 & 0.71 & 10.66 & 66.64  & 0.47 \\
                    &          & W &  & 1.86 & 0.66 & $-$62.15 & 0.53 & 0.50 & 29.23 & 17.10  & 0.48 \\
                    &          & D &  & 1.28 & 0.44 & $-$61.87 & 0.34 & 0.33 & 21.98 & 15.11  & 0.45 \\
         2014-02-28 & 56716 & K &  & 6.30 & 3.15 & $-$64.03 & 0.81 & 0.81 & 5.05  & 159.42 & 0.59 \\
                    &          & Q &  & 2.80 & 1.77 & $-$76.63 & 0.74 & 0.74 & 5.20  & 141.52 & 0.60 \\
                    &          & W &  & 1.80 & 0.71 & $-$61.59 & 0.56 & 0.55 & 19.06 & 28.89  & 0.58 \\
         2014-03-22 & 56740 & K &  & 5.37 & 2.99 & $-$75.45 & 0.85 & 0.85 & 4.98  & 170.71 & 0.58 \\
                    &          & Q &  & 2.62 & 1.56 & $-$74.61 & 0.80 & 0.79 & 10.86 & 73.15  & 0.72 \\
                    &          & W &  & 1.40 & 0.71 & $-$76.79 & 0.61 & 0.58 & 23.84 & 24.42  & 0.70 \\
                    &          & D &  & 0.96 & 0.48 & $-$72.7  & 0.32 & 0.31 & 21.50 & 14.49  & 0.71 \\
         2014-04-23 & 56771 & K &  & 5.31 & 3.03 & $-$78.69 & 0.86 & 0.86 & 4.77  & 180.07 & 0.64 \\
                    &          & Q &  & 2.63 & 1.54 & $-$77.77 & 0.90 & 0.90 & 7.09  & 126.72 & 0.80 \\
                    &          & W &  & 1.40 & 0.72 & $-$80.88 & 0.79 & 0.82 & 16.60 & 49.40  & 0.81 \\
                    &          & D &  & 0.97 & 0.48 & $-$74.73 & 0.58 & 0.57 & 22.06 & 25.86  & 0.79 \\
         2014-09-02 & 56903 & K &  & 5.23 & 3.04 & $-$82.95 & 1.08 & 1.08 & 2.27  & 474.39 & 0.44 \\
                    &          & Q &  & 2.59 & 1.53 & $-$79.77 & 1.12 & 1.12 & 9.39  & 119.12 & 0.44 \\
                    &          & W &  & 1.37 & 0.74 & $-$87.21 & 0.88 & 0.87 & 15.37 & 56.58  & 0.56 \\
                    &          & D &  & 1.36 & 0.74 & $-$85.89 & 0.92 & 0.91 & 40.35 & 22.66  & 0.51 \\
         2014-09-27 & 56928 & K &  & 5.33 & 2.98 & $-$80.54 & 1.04 & 1.03 & 11.60 & 89.12  & 0.67 \\
                    &          & W &  & 1.53 & 0.72 & $-$60.79 & 0.80 & 0.78 & 35.51 & 21.87  & 0.51 \\
         2014-10-29 & 56960 & K &  & 5.67 & 2.81 & $-$87.99 & 1.21 & 1.20 & 8.96  & 134.15 & 0.63 \\
                    &          & Q &  & 2.86 & 1.40 & $-$88.22 & 1.08 & 1.07 & 13.17 & 81.35  & 0.69 \\
                    &          & W &  & 1.47 & 0.68 & $-$87.72 & 0.75 & 0.73 & 19.07 & 38.19  & 0.62 \\
         2014-11-28 & 56989 & K &  & 5.71 & 2.82 & $-$75.66 & 1.54 & 1.53 & 21.04 & 72.54  & 0.73 \\
                    &          & Q &  & 2.79 & 1.46 & $-$71.32 & 1.58 & 1.57 & 9.41  & 167.27 & 0.64 \\
                    &          & W &  & 1.48 & 0.70 & $-$73.56 & 1.08 & 1.13 & 38.51 & 29.39  & 0.59 \\
         2014-12-26 & 57018 & K &  & 5.54 & 2.88 & $-$82.83 & 2.02 & 2.02 & 10.70 & 188.90 & 0.66 \\
         2015-01-15 & 57037 & K &  & 5.37 & 2.93 & $-$83.47 & 2.11 & 2.09 & 17.16 & 122.04 & 0.72 \\
                    &          & Q &  & 2.57 & 1.57 & $-$79.39 & 1.93 & 1.93 & 14.65 & 131.38 & 0.75 \\
                    &          & W &  & 1.38 & 0.71 & $-$81.48 & 1.38 & 1.36 & 36.28 & 37.51  & 0.93 \\
         2015-02-23 & 57076 & K &  & 5.37 & 2.95 & $-$86.16 & 2.59 & 2.59 & 8.18  & 317.11 & 0.51 \\
                    &          & Q &  & 2.60 & 1.54 & $-$87.84 & 2.47 & 2.47 & 11.54 & 213.79 & 0.56 \\
                    &          & W &  & 1.39 & 0.71 & $-$85.68 & 1.94 & 1.92 & 32.87 & 58.36  & 0.79 \\
         2015-03-26 & 57109 & K &  & 5.46 & 2.93 & $-$79.81 & 2.43 & 2.43 & 6.81  & 356.55 & 0.63 \\
                    &          & Q &  & 2.63 & 1.53 & $-$78.66 & 2.09 & 2.08 & 17.41 & 119.38 & 0.63 \\
                    &          & W &  & 1.39 & 0.72 & $-$81.74 & 1.38 & 1.34 & 33.44 & 40.03  & 0.70 \\
         2015-04-30 & 57144 & K &  & 6.47 & 2.67 & 79.66  & 1.94 & 1.94 & 25.24 & 77.01  & 0.61 \\
                    &          & W &  & 1.77 & 0.63 & 79.99  & 1.21 & 1.20 & 39.64 & 30.39  & 0.68 \\
                    &          & D &  & 1.10 & 0.44 & 88.57  & 1.13 & 1.10 & 125.81 & 8.76   & 0.48 \\
         2015-09-24 & 57290 & K &  & 5.61 & 2.95 & $-$70.86 & 1.92 & 1.92 & 16.46 & 116.62 & 0.90 \\
                    &          & Q &  & 3.04 & 1.45 & $-$68.40  & 1.66 & 1.65 & 12.37 & 133.21 & 0.68 \\
                    &          & W &  & 1.57 & 0.69 & $-$70.09 & 1.19 & 1.16 & 21.52 & 53.71  & 0.64 \\
         2015-10-23 & 57319 & K &  & 5.37 & 2.97 & $-$88.90  & 1.79 & 1.79 & 14.14 & 126.63 & 0.84 \\
                    &          & Q &  & 2.85 & 1.43 & $-$87.98 & 1.45 & 1.44 & 13.09 & 110.31 & 0.75 \\
                    &          & W &  & 1.50 & 0.69 & 86.09  & 1.03 & 1.01 & 38.81 & 25.98  & 0.66 \\
         2015-11-30 & 57357 & K &  & 5.17 & 3.13 & $-$85.41 & 1.79 & 1.79 & 11.53 & 155.14 & 0.87 \\
                    &          & Q &  & 2.59 & 1.56 & $-$84.21 & 1.68 & 1.68 & 9.74  & 172.09 & 0.63 \\
                    &          & W &  & 1.33 & 0.76 & $-$86.73 & 1.33 & 1.32 & 20.92 & 63.07  & 0.82 \\
                    &          & D &  & 0.88 & 0.52 & $-$83.01 & 0.87 & 0.87 & 41.42 & 20.97  & 0.75 \\
         2015-12-28 & 57385 & K &  & 5.31 & 3.08 & $-$85.92 & 1.81 & 1.81 & 13.77 & 131.24 & 0.95 \\
                    &          & Q &  & 2.65 & 1.56 & $-$86.71 & 1.82 & 1.81 & 7.39  & 245.50 & 0.80 \\
                    &          & W &  & 1.40 & 0.72 & $-$86.49 & 1.38 & 1.36 & 15.61 & 86.94  & 0.76 \\
                    &          & D &  & 0.91 & 0.45 & $-$81.19 & 0.72 & 0.72 & 11.33 & 63.41  & 0.81 \\
         2016-01-13 & 57401 & K &  & 5.26 & 3.11 & $-$88.32 & 1.69 & 1.69 & 9.20  & 183.72 & 0.69\\
                    &          & Q &  & 2.66 & 1.53 & 88.50   & 1.77 & 1.76 & 9.00 & 196.54  & 0.64 \\
                    &          & W &  & 1.40 & 0.73 & $-$89.84 & 1.55 & 1.52 & 23.02 & 67.22  & 0.78 \\
                    &          & D &  & 0.97 & 0.48 & 86.90   & 1.04 & 1.03 & 26.51 & 39.18  & 0.73 \\
         2016-03-01 & 57449 & K &  & 5.41 & 3.00 & $-$80.80  & 2.39 & 2.39 & 11.74 & 203.32 & 0.80 \\
                    &          & Q &  & 2.74 & 1.48 & $-$79.05 & 2.42 & 2.40 & 17.66 & 135.99 & 0.83 \\
                    &          & W &  & 1.38 & 0.73 & $-$81.56 & 1.84 & 1.82 & 22.69 & 80.23  & 0.84 \\
                    &          & D &  & 0.91 & 0.52 & $-$73.93 & 1.37 & 1.34 & 28.27 & 48.53  & 0.76 \\
         2016-04-24 & 57503 & K &  & 5.13 & 3.12 & $-$80.46 & 2.88 & 2.88 & 49.79 & 57.89  & 0.76 \\
                    &          & Q &  & 2.63 & 1.53 & $-$80.71 & 2.80 & 2.76 & 17.99 & 155.59  & 0.61 \\
                    &          & W &  & 1.33 & 0.76 & $-$85.64 & 1.89 & 1.87 & 27.17 & 68.82  & 0.75 \\
         2016-08-23 & 57624 & K &  & 6.44 & 3.05 & $-$57.94 & 1.38 & 1.38 & 16.30 & 84.83  & 0.54 \\
                    &          & D &  & 1.06 & 0.49 & $-$63.06 & 0.78 & 0.77 & 69.99 & 10.93  & 0.46 \\
         2016-10-18 & 57680 & K &  & 5.38 & 2.96 & $-$88.14 & 1.18 & 1.18 & 10.56 & 112.16 & 0.79 \\
                    &          & Q &  & 2.92 & 1.37 & $-$83.95 & 1.08 & 1.08 & 11.50 & 93.68  & 0.74 \\
                    &          & W &  & 1.39 & 0.72 & 88.08  & 0.87 & 0.87 & 19.39 & 44.82  & 0.75 \\
                    &          & D &  & 0.90 & 0.52 & 81.26  & 0.68 & 0.66 & 43.65 & 15.16  & 0.70 \\
         2016-11-27 & 57719 & K &  & 5.22 & 3.35 & 87.10   & 1.49 & 1.49 & 9.62  & 154.60 & 0.80 \\
                    &          & Q &  & 3.11 & 1.51 & 71.55  & 1.24 & 1.23 & 37.08 & 33.24  & 0.45 \\
                    &          & W &  & 1.56 & 0.77 & 68.18  & 1.14 & 1.13 & 29.35 & 38.39  & 0.48 \\
         2016-12-28 & 57750 & K &  & 5.12 & 3.14 & $-$85.72 & 1.95 & 1.95 & 12.93 & 150.86 & 0.75 \\
                    &          & Q &  & 2.68 & 1.50 & $-$88.74 & 1.71 & 1.70 & 20.09 & 84.69  & 0.84 \\
                    &          & W &  & 1.29 & 0.78 & $-$87.96 & 1.25 & 1.24 & 32.24 & 38.51  & 0.74 \\
                    &          & D &  & 0.81 & 0.59 & $-$86.55 & 0.79 & 0.79 & 31.10 & 25.36  & 0.74 \\
\end{longtable}
\tablefoot{
         (1) Date in Year-Month-Date,
         (2) Modified Julian Date,
         (3) observing frequency band,
         (4) major axis of beam,
         (5) minor axis of beam,
         (6) position angle of major axis,
         (7) CLEANed total flux density,
         (8) CLEANed peak flux density,
         (9) rms noise,
         (10) dynamic range,
         (11) image quality.
         }
}

\longtab[2]{
\begin{longtable}{cccccccc}
\caption{MODELFIT parameters from KVN Observations.\label{tbl2}}\\
\hline\hline
 &
 &
&
&
{$S_{\textup{total}}$} &
{$S_{\textup{peak}}$} &
{$d$} &
{$T_{\rm b}$} \\
Date & MJD & Band & & {(Jy)} & {(Jy/beam)} & {(mas)} & {($10^{10}$~K)} \\
(1)&(2)&(3)&&(4)&(5)&(6)&(7)\\
\hline
\endfirsthead
\caption{continued.}\\
\hline\hline
 &
 &
&
&
{$S_{\textup{total}}$} &
{$S_{\textup{peak}}$} &
{$d$} &
{$T_{\rm b}$} \\
Date & MJD & Band & & {(Jy)} & {(Jy/beam)} & {(mas)} & {($10^{10}$~K)} \\
(1)&(2)&(3)&&(4)&(5)&(6)&(7)\\
\hline
\endhead
\hline
\endfoot
         2012-12-05  &   56266   &   K   &   &  0.78 $\pm$ 0.01  &  0.76 $\pm$ 0.01  &  0.62 $\pm$0.01	&  0.93 $\pm$  0.03	\\
                     &           &   Q   &   &  0.82 $\pm$ 0.03  &  0.64 $\pm$ 0.02  &  1.12 $\pm$0.08	&  0.07 $\pm$  0.01	\\
         2013-01-16  &   56308   &   K   &   &  0.81 $\pm$ 0.00  &  0.78 $\pm$ 0.00  &  0.75 $\pm$0.00	&  0.65 $\pm$  0.01	\\
                     &           &   W   &   &  0.37 $\pm$ 0.02  &  0.25 $\pm$ 0.01  &  0.72 $\pm$0.07	&  0.02 $\pm$  0.00	\\
                     &           &   D   &   &  0.33 $\pm$ 0.00  &  0.27 $\pm$ 0.00  &  0.51 $\pm$0.03	&  0.02 $\pm$  0.00	\\
         2013-02-27  &   56349   &   K   &   &  0.76 $\pm$ 0.01  &  0.72 $\pm$ 0.01  &  0.93 $\pm$0.02	&  0.39 $\pm$  0.02	\\
                     &           &   Q   &   &  0.80 $\pm$ 0.02  &  0.69 $\pm$ 0.01  &  0.76 $\pm$0.03	&  0.16 $\pm$  0.01	\\
                     &           &   W   &   &  0.68 $\pm$ 0.03  &  0.53 $\pm$ 0.02  &  0.52 $\pm$0.03	&  0.07 $\pm$  0.01	\\
                     &           &   D   &   &  0.32 $\pm$ 0.05  &  0.35 $\pm$ 0.04  &  0.23 $\pm$0.11	&  0.08 $\pm$  0.09 \\
         2013-03-28  &   56381   &   K   &   &  0.76 $\pm$ 0.01  &  0.73 $\pm$ 0.00  &  0.80 $\pm$0.01	&  0.53 $\pm$  0.01	\\
                     &           &   Q   &   &  0.82 $\pm$ 0.02  &  0.72 $\pm$ 0.01  &  0.72 $\pm$0.03	&  0.18 $\pm$  0.02	\\
                     &           &   W   &   &  0.86 $\pm$ 0.04  &  0.58 $\pm$ 0.02  &  0.67 $\pm$0.06	&  0.06 $\pm$  0.01	\\
                     &           &   D   &   &  1.07 $\pm$ 0.07  &  0.57 $\pm$ 0.03  &  0.62 $\pm$0.12	&  0.04 $\pm$  0.02	\\
         2013-04-11  &   56394   &   K   &   &  0.89 $\pm$ 0.01  &  0.87 $\pm$ 0.01  &  0.51 $\pm$0.00	&  1.51 $\pm$  0.03	\\
                     &           &   Q   &   &  0.92 $\pm$ 0.02  &  0.87 $\pm$ 0.01  &  0.50 $\pm$0.01	&  0.41 $\pm$  0.02	\\
                     &           &   W   &   &  0.95 $\pm$ 0.07  &  0.59 $\pm$ 0.04  &  0.40 $\pm$0.01	&  0.15 $\pm$  0.01	\\
         2013-05-08  &   56422   &   K   &   &  0.91 $\pm$ 0.01  &  0.86 $\pm$ 0.01  &  0.94 $\pm$0.02	&  0.47 $\pm$  0.02	\\
                     &           &   W   &   &  0.84 $\pm$ 0.08  &  0.49 $\pm$ 0.04  &  0.80 $\pm$0.15	&  0.04 $\pm$  0.02	\\
         2013-09-24  &   56559   &   K   &   &  0.83 $\pm$ 0.03  &  0.80 $\pm$ 0.02  &  0.70 $\pm$0.01	&  0.75 $\pm$  0.03	\\
                     &           &   Q   &   &  0.71 $\pm$ 0.01  &  0.70 $\pm$ 0.01  &  0.33 $\pm$0.04	&  0.72 $\pm$  0.02	\\
         2013-10-15  &   56581   &   K   &   &  0.80 $\pm$ 0.01  &  0.80 $\pm$ 0.01  &  $<$1.15 		&  $>$0.27 \\
                     &           &   Q   &   &  0.75 $\pm$ 0.02  &  0.74 $\pm$ 0.02  &  0.30 $\pm$0.00	&  0.94 $\pm$  0.03	\\
                     &           &   D   &   &  0.78 $\pm$ 0.06  &  0.57 $\pm$ 0.03  &  0.59 $\pm$0.06	&  0.03 $\pm$  0.01	\\
         2013-11-20  &   56616   &   K   &   &  0.87 $\pm$ 0.03  &  0.85 $\pm$ 0.02  &  0.52 $\pm$0.01	&  1.42 $\pm$  0.05   \\
                     &           &   Q   &   &  0.88 $\pm$ 0.02  &  0.80 $\pm$ 0.01  &  0.61 $\pm$0.02	&  0.26 $\pm$  0.02	\\
                     &           &   W   &   &  0.85 $\pm$ 0.01  &  0.71 $\pm$ 0.01  &  0.41 $\pm$0.01	&  0.14 $\pm$  0.01	\\
                     &           &   D   &   &  0.66 $\pm$ 0.03  &  0.57 $\pm$ 0.02  &  0.25 $\pm$0.01	&  0.13 $\pm$  0.02	\\
         2013-12-24  &   56651   &   K   &   &  0.83 $\pm$ 0.01  &  0.80 $\pm$ 0.01  &  0.67 $\pm$0.01	&  0.83 $\pm$  0.02	\\
                     &           &   Q   &   &  0.77 $\pm$ 0.01  &  0.71 $\pm$ 0.01  &  0.58 $\pm$0.01	&  0.25 $\pm$  0.01	\\
                     &           &   W   &   &  0.70 $\pm$ 0.02  &  0.51 $\pm$ 0.01  &  0.56 $\pm$0.03	&  0.06 $\pm$  0.01	\\
                     &           &   D   &   &  0.44 $\pm$ 0.03  &  0.34 $\pm$ 0.02  &  0.33 $\pm$0.01	&  0.05 $\pm$  0.01	\\
         2014-02-28  &   56716   &   K   &   &  0.80 $\pm$ 0.01  &  0.81 $\pm$ 0.01  &  $<$1.37 		&  $>$0.19 \\
                     &           &   Q   &   &  0.76 $\pm$ 0.01  &  0.73 $\pm$ 0.01  &  0.40 $\pm$0.00	&  0.54 $\pm$  0.01	\\
                     &           &   W   &   &  0.61 $\pm$ 0.03  &  0.55 $\pm$ 0.02  &  0.33 $\pm$0.01	&  0.16 $\pm$  0.01	\\
         2014-03-22  &   56740   &   K   &   &  0.86 $\pm$ 0.01  &  0.85 $\pm$ 0.01  &  0.38 $\pm$0.00	&  2.64 $\pm$  0.05	\\
                     &           &   Q   &   &  0.88 $\pm$ 0.02  &  0.79 $\pm$ 0.01  &  0.64 $\pm$0.02	&  0.24 $\pm$  0.02	\\
                     &           &   W   &   &  0.81 $\pm$ 0.03  &  0.59 $\pm$ 0.02  &  0.56 $\pm$0.03	&  0.07 $\pm$  0.01	\\
                     &           &   D   &   &  0.35 $\pm$ 0.03  &  0.31 $\pm$ 0.02  &  0.23 $\pm$0.02	&  0.09 $\pm$  0.01	\\
         2014-04-23  &   56771   &   K   &   &  0.85 $\pm$ 0.01  &  0.86 $\pm$ 0.01  &  $<$1.23 		&  $>$0.25 \\
                     &           &   Q   &   &  0.93 $\pm$ 0.01  &  0.90 $\pm$ 0.01  &  0.38 $\pm$0.01	&  0.72 $\pm$  0.04	\\
                     &           &   W   &   &  0.89 $\pm$ 0.02  &  0.82 $\pm$ 0.02  &  0.28 $\pm$0.01	&  0.31 $\pm$  0.02	\\
                     &           &   D   &   &  0.61 $\pm$ 0.03  &  0.57 $\pm$ 0.02  &  0.16 $\pm$0.00	&  0.30 $\pm$  0.01	\\
         2014-09-02  &   56903   &   K   &   &  1.07 $\pm$ 0.01  &  1.08 $\pm$ 0.01  &  $<$1.37     	&  $>$0.26	\\
                     &           &   Q   &   &  1.11 $\pm$ 0.02  &  1.12 $\pm$ 0.01  &  $<$0.33 		&  $>$1.18 \\
                     &           &   W   &   &  0.98 $\pm$ 0.01  &  0.86 $\pm$ 0.01  &  0.34 $\pm$0.00	&  0.24 $\pm$  0.01	\\
                     &           &   D   &   &  1.20 $\pm$ 0.01  &  0.92 $\pm$ 0.01  &  0.52 $\pm$0.01	&  0.06 $\pm$  0.00	\\
         2014-09-27  &   56928   &   K   &   &  1.08 $\pm$ 0.02  &  1.03 $\pm$ 0.01  &  0.79 $\pm$0.01	&  0.77 $\pm$  0.01	\\
                     &           &   W   &   &  0.90 $\pm$ 0.06  &  0.77 $\pm$ 0.04  &  0.38 $\pm$0.03	&  0.17 $\pm$  0.03	\\
         2014-10-29  &   56960   &   K   &   &  1.27 $\pm$ 0.01  &  1.20 $\pm$ 0.01  &  0.88 $\pm$0.01	&  0.74 $\pm$  0.02	\\
                     &           &   Q   &   &  1.13 $\pm$ 0.02  &  1.07 $\pm$ 0.01  &  0.42 $\pm$0.01	&  0.70 $\pm$  0.05	\\
                     &           &   W   &   &  0.84 $\pm$ 0.03  &  0.72 $\pm$ 0.02  &  0.35 $\pm$0.01	&  0.19 $\pm$  0.02	\\
         2014-11-28  &   56989   &   K   &   &  1.65 $\pm$ 0.03  &  1.52 $\pm$ 0.02  &  1.07 $\pm$0.04	&  0.64 $\pm$  0.05	\\
                     &           &   Q   &   &  1.59 $\pm$ 0.01  &  1.57 $\pm$ 0.01  &  0.20 $\pm$0.00	&  4.35 $\pm$  0.09	\\
                     &           &   W   &   &  1.35 $\pm$ 0.04  &  1.13 $\pm$ 0.03  &  0.40 $\pm$0.04	&  0.24 $\pm$  0.05	\\
         2014-12-26  &   57018   &   K   &   &  2.03 $\pm$ 0.02  &  2.02 $\pm$ 0.01  &  0.26 $\pm$0.00	&  13.17 $\pm$ 0.14 \\
         2015-01-15  &   57037   &   K   &   &  2.13 $\pm$ 0.03  &  2.09 $\pm$ 0.02  &  0.51 $\pm$0.01	&  3.73 $\pm$  0.12	\\
                     &           &   Q   &   &  1.94 $\pm$ 0.02  &  1.92 $\pm$ 0.02  &  0.18 $\pm$0.00	&  6.56 $\pm$  0.19	\\
                     &           &   W   &   &  1.52 $\pm$ 0.05  &  1.35 $\pm$ 0.04  &  0.32 $\pm$0.01	&  0.42 $\pm$  0.03	\\
         2015-02-23  &   57076   &   K   &   &  2.58 $\pm$ 0.01  &  2.59 $\pm$ 0.01  &  $<$1.10	&  $>$0.96	 \\
                     &           &   Q   &   &  2.54 $\pm$ 0.01  &  2.46 $\pm$ 0.01  &  0.35 $\pm$0.00	&  2.36 $\pm$  0.03	 \\
                     &           &   W   &   &  2.03 $\pm$ 0.05  &  1.91 $\pm$ 0.03  &  0.23 $\pm$0.01	&  1.07 $\pm$  0.07	 \\
         2015-03-26  &   57109   &   K   &   &  2.43 $\pm$ 0.01  &  2.43 $\pm$ 0.01  &  $<$0.21 		&  $>$26.04 \\
                     &           &   Q   &   &  2.20 $\pm$ 0.02  &  2.07 $\pm$ 0.02  &  0.48 $\pm$0.01	&  1.06 $\pm$  0.04	 \\
                     &           &   W   &   &  1.68 $\pm$ 0.04  &  1.34 $\pm$ 0.03  &  0.46 $\pm$0.04	&  0.23 $\pm$  0.05	 \\
         2015-04-30  &   57144   &   K   &   &  1.94 $\pm$ 0.04  &  1.94 $\pm$ 0.03  &  $<$1.27 		&  $>$0.54 \\
                     &           &   W   &   &  1.20 $\pm$ 0.06  &  1.20 $\pm$ 0.04  &  $<$0.07 		&  $>$7.27 \\
                     &           &   D   &   &  2.15 $\pm$ 0.12  &  1.26 $\pm$ 0.06  &  0.52 $\pm$0.04	&  0.10 $\pm$  0.02	 \\
         2015-09-24  &   57290   &   K   &   &  1.92 $\pm$ 0.02  &  1.92 $\pm$ 0.02  &  $<$0.92 		&  $>$1.02 \\
                     &           &   Q   &   &  1.73 $\pm$ 0.02  &  1.64 $\pm$ 0.01  &  0.42 $\pm$0.01	&  1.08 $\pm$  0.04	 \\
                     &           &   W   &   &  1.36 $\pm$ 0.03  &  1.15 $\pm$ 0.02  &  0.39 $\pm$0.01	&  0.25 $\pm$  0.02	 \\
         2015-10-23  &   57319   &   K   &   &  1.78 $\pm$ 0.02  &  1.79 $\pm$ 0.02  &  $<$1.17 		&  $>$0.58 \\
                     &           &   Q   &   &  1.53 $\pm$ 0.02  &  1.44 $\pm$ 0.01  &  0.45 $\pm$0.01	&  0.84 $\pm$  0.05	 \\
                     &           &   W   &   &  1.30 $\pm$ 0.05  &  1.01 $\pm$ 0.03  &  0.46 $\pm$0.06	&  0.17 $\pm$  0.05	 \\
         2015-11-30  &   57357   &   K   &   &  1.78 $\pm$ 0.02  &  1.79 $\pm$ 0.01  &  $<$1.22 		&  $>$0.54 \\
                     &           &   Q   &   &  1.74 $\pm$ 0.01  &  1.67 $\pm$ 0.01  &  0.38 $\pm$0.00	&  1.37 $\pm$  0.03	 \\
                     &           &   W   &   &  1.59 $\pm$ 0.04  &  1.31 $\pm$ 0.02  &  0.39 $\pm$0.08	&  0.30 $\pm$  0.15	 \\
                     &           &   D   &   &  0.84 $\pm$ 0.06  &  0.86 $\pm$ 0.04  &  0.12 $\pm$0.03	&  0.72 $\pm$  0.45    \\
         2015-12-28  &   57385   &   K   &   &  1.83 $\pm$ 0.02  &  1.80 $\pm$ 0.02  &  $<$0.98	        &  $>$0.86	 \\
                     &           &   Q   &   &  1.89 $\pm$ 0.01  &  1.81 $\pm$ 0.01  &  0.41 $\pm$0.01	&  1.26 $\pm$  0.04	 \\
                     &           &   W   &   &  1.49 $\pm$ 0.03  &  1.35 $\pm$ 0.02  &  $<$0.21	        &  $>$0.93	 \\
                     &           &   D   &   &  0.74 $\pm$ 0.02  &  0.72 $\pm$ 0.01  &  0.11 $\pm$0.00	&  0.85 $\pm$  0.01	 \\
         2016-01-13  &   57401   &   K   &   &  1.68 $\pm$ 0.02  &  1.69 $\pm$ 0.01  &  $<$1.39        	&  $>$0.39	 \\
                     &           &   Q   &   &  1.89 $\pm$ 0.03  &  1.55 $\pm$ 0.02  &  $<$0.28	        &  $>$2.57	 \\
                     &           &   W   &   &  1.90 $\pm$ 0.07  &  1.21 $\pm$ 0.04  &  $<$0.22     	&  $>$0.86	 \\
                     &           &   D   &   &  1.23 $\pm$ 0.07  &  0.79 $\pm$ 0.04  &  0.04 $\pm$0.00	&  8.77 $\pm$  1.67	 \\
         2016-03-01  &   57449   &   K   &   &  2.35 $\pm$ 0.02  &  2.38 $\pm$ 0.02  &  $<$1.61	        &  $>$0.40	 \\
                     &           &   Q   &   &  2.65 $\pm$ 0.02  &  2.38 $\pm$ 0.01  &  0.62 $\pm$0.01	&  0.78 $\pm$  0.04	 \\
                     &           &   W   &   &  2.08 $\pm$ 0.03  &  1.81 $\pm$ 0.02  &  0.36 $\pm$0.00	&  0.46 $\pm$  0.01	 \\
                     &           &   D   &   &  1.29 $\pm$ 0.05  &  1.24 $\pm$ 0.03  &  $<$0.04	        &  $>$9.41	   \\
         2016-04-24  &   57503   &   K   &   &  2.92 $\pm$ 0.07  &  2.88 $\pm$ 0.05  &  $<$1.02	        &  $>$1.25	   \\
                     &           &   Q   &   &  2.99 $\pm$ 0.04  &  2.62 $\pm$ 0.03  &  0.39 $\pm$0.00	&  2.19 $\pm$  0.04	   \\
                     &           &   W   &   &  2.01 $\pm$ 0.04  &  1.86 $\pm$ 0.03  &  0.24 $\pm$0.03	&  0.95 $\pm$  0.25	   \\
         2016-08-23  &   57624   &   K   &   &  1.37 $\pm$ 0.02  &  1.38 $\pm$ 0.02  &  $<$1.60 		&  $>$0.24 \\
                     &           &   D   &   &  0.87 $\pm$ 0.11  &  0.76 $\pm$ 0.07  &  0.24 $\pm$0.02	&  0.19 $\pm$  0.04	   \\
         2016-10-18  &   57680   &   K   &   &  1.19 $\pm$ 0.02  &  1.19 $\pm$ 0.01  &  $<$0.81 		&  $>$0.82 \\
                     &           &   Q   &   &  1.18 $\pm$ 0.01  &  1.07 $\pm$ 0.01  &  0.58 $\pm$0.01	&  0.40 $\pm$  0.02	   \\
                     &           &   W   &   &  1.02 $\pm$ 0.04  &  0.86 $\pm$ 0.02  &  0.32 $\pm$0.08	&  0.28 $\pm$  0.16	   \\
                     &           &   D   &   &  0.87 $\pm$ 0.06  &  0.66 $\pm$ 0.04  &  0.36 $\pm$0.03	&  0.08 $\pm$  0.02	   \\
         2016-11-27  &   57719   &   K   &   &  1.49 $\pm$ 0.01  &  1.49 $\pm$ 0.01  &  $<$1.07 		&  $>$0.58 \\
                     &           &   Q   &   &  1.44 $\pm$ 0.02  &  1.23 $\pm$ 0.01  &  0.81 $\pm$0.03	&  0.24 $\pm$  0.02	   \\
                     &           &   W   &   &  1.31 $\pm$ 0.07  &  1.12 $\pm$ 0.05  &  0.36 $\pm$0.06	&  0.28 $\pm$  0.12	   \\
         2016-12-28  &   57750   &   K   &   &  1.95 $\pm$ 0.02  &  1.95 $\pm$ 0.01  &  $<$1.22	&  $>$0.59	   \\
                     &           &   Q   &   &  1.92 $\pm$ 0.02  &  1.69 $\pm$ 0.01  &  0.69 $\pm$0.03	&  0.46 $\pm$  0.04	   \\
                     &           &   W   &   &  1.50 $\pm$ 0.06  &  1.23 $\pm$ 0.04  &  0.40 $\pm$0.08	&  0.26 $\pm$  0.12	   \\
                     &           &   D   &   &  0.78 $\pm$ 0.04  &  0.79 $\pm$ 0.03  &  0.08 $\pm$0.01	&  1.46 $\pm$  0.61 \\

\end{longtable}
\tablefoot{
(1) Date in Year-Month-Day,
(2) Modified Julian Date,
(3) observing frequency band (K, Q, W, and D for 22, 43, 86, and 129~GHz, respectively),
(4) observed total flux density in Jy,
(5) observed peak flux of core components in Jy,
(6) observed core size in mas including its upper limit with a symbol of '$<$',
(7) estimated brightness temperature in $10^{10}$K including its lower limit with a symbol of '$>$'.
The uncertainties of 0.00 are less than 0.005.
}}


\section{Brightness Temperature}\label{appb}
The definition of brightness temperature is as follows:
\begin{equation}
T_b = \frac{h \nu}{k {\rm ln} \left(1+\frac{2h\nu^3}{I_\nu c^2}\right)} .
\end{equation}
Under the Rayleigh-Jeans approximation, this simplifies to
\begin{equation}
T_b = \frac{I_\nu c^2}{2k\nu^2}.    
\end{equation}
The intensity $I_\nu$ can be defined as ${S}/{\Omega}$, where $\Omega$ is ${\pi \theta^2}/{\rm 4ln2}$ in circular Gaussian structure.
Therefore we can write the equation as 
\begin{equation}
T_b = \frac{c^2}{2 k \nu^2 \Omega}S = \frac{2\rm ln2}{\pi k } \frac{c^2S}{\nu^2 \theta^2}. \end{equation}
In order to estimate the brightness temperature assuming that the emission is dominated by variability emission region,
we can estimate variability size as 
\begin{equation}
\theta_{\rm var} = \frac{c\Delta t} {D_{\rm A}(1+z)}.     
\end{equation}
If we replace $\theta$ as $\theta_{\rm var} $, we can obtain 
\begin{equation}
T_b = \frac{2 {\rm ln}2}{\pi k} \frac{D_{\rm A}^2 (1+z)^2 S}{\nu^2 \Delta t^2}.
\end{equation}
Here, $\nu$ is frequency in the rest frame.
In order to convert to the observer's frame by referring $\nu = \nu_{\rm obs}(1+z)$ and $D_{\rm A}={D_{\rm L}}/{(1+z)^2}$, we can obtain
\begin{equation}
T_b = \frac{2 {\rm ln}2}{\pi k} \frac{D_{\rm L}^2 S}{\nu_{\rm obs}^2 \Delta t^2 (1+z)^4}.
\end{equation}
Computing the numerical factor, 
we obtain a final, convenient expression of 
\begin{equation}
T_{\rm B}^{\rm var}=4.077\times10^{13}    
\left(\frac{D_{\rm L}}{[\rm Mpc]}\right)^2 
\left(\frac{\nu_{\rm obs}}{[\rm GHz]}\right)^{-2}
\left(\frac{\Delta t}{[\rm day]}\right)^{-2}
\frac{S}{(1+z)^4}.
\end{equation}

\begin{table}[h!]
\centering
\caption{Computed numerical factors { between two different sets of units for brightness temperature estimation.}\label{tbl12}}
\begin{tabular}{ccccc}
\tableline
\tableline
  S  &  $D_{\rm L}$ &  $\nu$ & $\Delta t$ &  Numerical factor \\
\tableline
  [Jy]  & [m]   & [Hz]  & [s]     & $3.195\times10^{-4}$ \\
  
  [Jy]  & [Mpc] & [GHz] & [yr]   & $3.06\times10^{8}$ \\
\tableline
\end{tabular}
\end{table}

\section{Error Propagation}\label{appc}
We estimated the uncertainty of magnetic field strength using basic error propagation.
When the magnetic field strength $B$ is estimated by $A^b$ (i.e. $B \propto A^b$), the uncertainty, $\sigma_B$, can be calculated by $\sigma_B = Bb\frac{\sigma_A}{A}$.
Therefore, the uncertainty of magnetic field strength in SSA, $\sigma_{B_{\rm SSA}}$, is proportional to
$4\frac{\sigma_{\theta_{\rm r}}}{\theta_{\rm r}}$ and $5\frac{\sigma_{\nu_{\rm r}}}{\nu_{\rm r}}$, and 
the uncertainty of magnetic field strength in equipartition, $\sigma_{B_{\rm eq}}$, is proportional to
$-\frac{6}{7}\frac{\sigma_{\theta_{\rm r}}}{\theta_{\rm r}}$ and $\frac{1}{7}\frac{\sigma_{\nu_{\rm r}}}{\nu_{\rm r}}$.
Moreover, the lower uncertainty of $B_{\rm SSA}$ is larger than $B_{\rm SSA}$ itself (e.g. $B_{\rm SSA}$ < $\sigma_{B_{\rm SSA}}$).
Therefore we compute the lower uncertainty of $B_{\rm SSA}$ in logarithm as follows:
\begin{eqnarray}
    \sigma_f &= &\left|{\frac{\sigma_{B_{\rm SSA}}}{{\rm ln}(10)B_{\rm SSA}}}\right|, \\
    \sigma_{B_{\rm SSA,low}} &= &B_{\rm SSA} - 10^{-\sigma_{f}}B_{\rm SSA}.
\end{eqnarray}
\end{appendix}

\end{document}